\documentclass[graybox]{svmult}

\pdfoutput=1
\usepackage{type1cm}
\usepackage{makeidx}         % allows index generation

\usepackage[pdftex]{graphicx} % For including figures
\graphicspath{{./Figures/}}

\usepackage{multicol}        % used for the two-column index
\usepackage[bottom]{footmisc}% places footnotes at page bottom

\usepackage{newtxtext}       % 
\usepackage[varvw]{newtxmath}       % selects Times Roman as basic font

\usepackage{epstopdf} % For using eps files in figures
\usepackage{float} % For being able to make environments float
\usepackage{subcaption} % For subfigures
\usepackage{amsmath, dsfont} % For various mathematical commands and symbols
\usepackage{enumitem}

\usepackage{url}
\usepackage{hyperref}
\hypersetup{%
    colorlinks=true,
    linkcolor={blue!80},
    citecolor={blue!80},
    urlcolor={black}
}

\newcommand{\T}{^\top} % Transpose
\newcommand{\cc}[1]{{\mathcal{#1}}} % Short for giving it the font for sets
\def\vv#1{{ \rm \bf{#1}}} % Bold
\def\I{\mathbf{I}} % Identity matrix
\def\R{\mathbb{R}} % Set of real numbers
\def\N{\mathbb{N}} % Set of natural numbers
\newcommand{\ones}[1]{\mathds{1}_{#1}} % Vector of ones of dimension #1

\def\set#1#2{\left\{  #1 : #2 \right\}}
\def\Sum#1#2{\sum\limits_{#1}^{#2}} % Sum with limits
\def\st{{\rm s.t.}} % Text for "subject to"
\def\diag{\texttt{diag}} % For diagonal matrices

\def\nx{{n_x}}
\def\nu{{n_u}}
\def\ny{{n_y}}
\def\xr{x_r} % State reference
\def\ur{u_r} % Input reference

\def\xs{x_s} % Artificial state reference
\def\us{u_s} % Artificial input reference
\def\xo{x_r^\circ} % Optimal reachable state reference
\def\uo{u_r^\circ} % Optimal reachable input reference
\def\Vo{V_o} % Symbol for the offset cost function

\def\Tp{T_p} % Period
\def\Vp{V_p} % Symbol for the offset cost function of perMPCT

\def\cD{\cc{D}} % Set for dynamics of a harmonic signal
\def\cC{\cc{C}_\sigma} % Set for constraints of a harmonic signal

\newcommand{\yLB}{\underline{y}}
\newcommand{\yLBj}{\underline{y}_{(i)}}
\newcommand{\yUB}{\overline{y}}
\newcommand{\yUBj}{\overline{y}_{(i)}}
\def\xe{x_e} % Notation for x_e
\def\xs{x_s} % Notation for x_s
\def\xc{x_c} % Notation for x_c
\def\ue{u_e} % Notation for u_e
\def\us{u_s} % Notation for u_s
\def\uc{u_c} % Notation for u_c
\def\xh{x_h} % Notation for the sequence of x_{h,j}
\def\uh{u_h} % Notation for the sequence of u_{h,j}
\def\xH{\vv{x}_h} % Notation for the shorthand for (x_e, x_s, x_c)
\def\uH{\vv{u}_h} % Notation for the shorthand for (u_e, u_si, u_ci)
\def\VfHMPC{V_h} % Notation for the offset cost of the HMPC controller
\def\xHo{\vv{x}_h^\circ} % Notation for the shorthand for (x_e^\circ, x_s^\circ, x_c^\circ)
\def\uHo{\vv{u}_h^\circ} % Notation for the shorthand for (u_e^\circ, u_s^\circ, u_c^\circ)
\def\xeo{x_e^\circ} % Notation for x_e
\def\ueo{u_e^\circ} % Notation for u_e

\def\pBP{p} % Position of the ball
\def\aBP{\theta} % Angle of the plate

\makeindex  

\begin{document}
% Fakesection Frontmater

\title*{Recent advancements on MPC for tracking: periodic and harmonic formulations}

\author{Pablo~Krupa\orcidID{0000-0002-6238-1166} and \\ Daniel~Limon\orcidID{0000-0001-9334-7289} and\\ Teodoro~Alamo\orcidID{0000-0002-0623-8146}}
\institute{Pablo~Krupa,~Daniel~Limon,~Teodoro~Alamo \at Universidad de Sevilla, Seville, Spain, \email{pkrupa@us.es, dlm@us.es, talamo@us.es}
}
\institute{Pablo~Krupa \at Gran Sasso Science Institute (GSSI), L'Aquila, Italy, \email{pablo.krupa@gssi.it}
    \and Daniel~Limon,~Teodoro~Alamo \at Universidad de Sevilla, Seville, Spain, \email{dlm@us.es, talamo@us.es}
}
\maketitle

\abstract*{
The main benefit of model predictive control (MPC) is its ability to steer the system to a given reference without violating the constraints while minimizing some objective.
Furthermore, a suitably designed MPC controller guarantees asymptotic stability of the closed-loop system to the given reference as long as its optimization problem is feasible at the initial state of the system.
Therefore, one of the limitations of classical MPC is that changing the reference may lead to an unfeasible MPC problem.
Furthermore, due to a lack of deep knowledge of the system, it is possible for the user to provide a desired reference that is unfeasible or non-attainable for the MPC controller, leading to the same problem.
This chapter summarizes MPC formulations recently proposed that have been designed to address these issues.
In particular, thanks to the addition of an artificial reference as decision variable, the formulations achieve asymptotic stability and recursive feasibility guarantees regardless of the reference provided by the user, even if it is changed online or if it violates the system constraints.
We show a recent formulation which extends this idea, achieving better performance and larger domains of attraction when working with small prediction horizons.
Additional benefits of these formulations, when compared to classical MPC, are also discussed and highlighted with illustrative examples.
}

\abstract{
The main benefit of model predictive control (MPC) is its ability to steer the system to a given reference without violating the constraints while minimizing some objective.
Furthermore, a suitably designed MPC controller guarantees asymptotic stability of the closed-loop system to the given reference as long as its optimization problem is feasible at the initial state of the system.
Therefore, one of the limitations of classical MPC is that changing the reference may lead to an unfeasible MPC problem.
Furthermore, due to a lack of deep knowledge of the system, it is possible for the user to provide a desired reference that is unfeasible or non-attainable for the MPC controller, leading to the same problem.
This chapter summarizes MPC formulations recently proposed that have been designed to address these issues.
In particular, thanks to the addition of an artificial reference as decision variable, the formulations achieve asymptotic stability and recursive feasibility guarantees regardless of the reference provided by the user, even if it is changed online or if it violates the system constraints.
We show a recent formulation which extends this idea, achieving better performance and larger domains of attraction when working with small prediction horizons.
Additional benefits of these formulations, when compared to classical MPC, are also discussed and highlighted with illustrative~examples.
}

% Fakesection Notation
\vspace*{2em}
\noindent\textbf{Notation: }
We denote the transpose of $x \in \R^n$ or $A \in \R^{m \times p}$ by $x\T$ and $A\T$.
Given $x \in \R^n$ and a positive definite $Q \in \R^{n \times n}$, we denote $\| x \|_Q \doteq \sqrt{x\T Q x}$ and $\| x \| \doteq \sqrt{x\T x}$.
The set of natural numbers (including $0$) is denoted by $\N$.
Given $i, j \in \N$ with $i \leq j$, $\N_i^j \doteq \{i, i+1, \dots, j\}$.
For two vectors $x, y \in \R^n$, $x \leq (<) \, y$ denotes component-wise inequalities.
The vector of ones of dimension $n$ is denoted by $\ones{n}$.
Element $i \in \N_1^n$ of a vector $x \in \R^n$ is denoted by $x_{(i)}$.
$\diag(\alpha_1, \dots \alpha_n) \in \R^{n \times n}$ is the diagonal matrix with entries $\alpha_i \in \R$, $i \in \N_1^n$.

\section{Introduction} \label{sec:intro}

Throughout this chapter, we consider the problem of controlling a system described by a controllable linear time-invariant state-space model
\begin{equation} \label{eq:model}
    x(t+1) = A x(t) + B u(t),
\end{equation}
where $x(t) \in \R^\nx$ and $u(t) \in \R^\nu$ are the state and control input at the discrete time instant $t$, respectively.
We also consider that the system is subject to coupled input-output constraints
\begin{equation} \label{eq:model:constraints}
    \yLB \leq E x(t) + F u(t) \leq \yUB,
\end{equation}
where we assume that the bounds $\yLB, \yUB \in \R^\ny$ satisfy $\yLB < \yUB$.

Our control objective is to steer~\eqref{eq:model} to a \emph{desired reference} $(\xr, \ur) \in \R^\nx \times \R^\nu$ while satisfying the constraints~\eqref{eq:model:constraints}.
Model Predictive Control (MPC) \cite{RMD17, Camacho_S_2013} is an optimization-based control technique that is particularly suitable for this task.
At each sample time, MPC solves a finite-horizon optimal control problem using the prediction model~\eqref{eq:model} and the current measurement (or estimate) of the system state $x(t)$.
The constraints of the system are explicitly considered in the optimal control problem.
Thus, under nominal conditions, a suitably designed MPC controller will steer the initial state $x(0)$ to the desired reference $\xr$ while satisfying the system constraints \cite{RMD17}, as long as the desired reference is an \emph{admissible steady state} of the system.

\begin{definition}[Admissible steady state] \label{def:ass}
    A pair $(x, u) \in \R^\nx \times \R^\nu$ is said to be an \emph{admissible steady state} of system~\eqref{eq:model} subject to~\eqref{eq:model:constraints} if it satisfies $x = A x + B u$ and $\yLB \leq E x + F u \leq \yUB$. 
    Furthermore, we say that it is \emph{strictly admissible} if it strictly satisfies the previous inequality, i.e., $\yLB < E x + F u < \yUB$.
\end{definition}

Let us start by considering a standard MPC with terminal equality constraint:
\begin{subequations} \label{eq:equMPC}
\begin{align}  
    \min\limits_{\vv{x}, \vv{u}} \;& \Sum{k = 0}{N-1} \| x(k|t) - \xr \|^2_Q + \| u(k|t) - \ur \|^2_R  \label{eq:equMPC:cost_function}\\
    \st & \; x(0|t) = x(t), \label{eq:equMPC:initial} \\
        & \; x(k+1|t) = A x(k|t) + B u(k|t), \; k\in\N_0^{N-1}, \label{eq:equMPC:prediction} \\
        & \; \yLB \leq E x(k|t) + F u(k|t) \leq \yUB, \; k\in\N_0^{N-1}, \label{eq:equMPC:constraints} \\
        & \; x(N|t) = \xr, \label{eq:equMPC:terminal}
\end{align}
\end{subequations}
where $\vv{x} = (x(0|t), x(1|t), \dots x(N|t))$ and $\vv{u} = (u(0|t), u(1|t), \dots, u(N-1|t))$ are the predicted states and inputs throughout the prediction horizon $N \in \N$; and the cost function matrices $Q \in \R^{\nx \times \nx}$ and $R \in \R^{\nu \times \nu}$ are assumed to be positive definite.
The use of the terminal equality constraint $x(N|t) = \xr$ provides the MPC controller~\eqref{eq:equMPC} with stability guarantees under rather mild assumption \cite{RMD17,Camacho_S_2013,Limon_A_2008}.
The classical alternative to using a terminal equality constraint is to use a \emph{terminal set}, where the terminal state $x(N|t)$ is forced to lie within a suitable invariant set of the system.
The use of a terminal equality constraint has the advantage of not requiring the computation of the terminal set, which can be challenging to obtain for non-trivial systems \cite{Blanchini_A_1999}.
Finally, it leads to a simple Quadratic Programming (QP) problem that can be solved online using one of the many available efficient QP solvers \cite{Stellato_OSQP, ODonoghue_SCS_21, Krupa_TCST_20, Spcies}.
The control law of~\eqref{eq:equMPC} is $u(t) = u^*(0|t)$, where $u^*(0|t)$ is the value of $u(0|t)$ corresponding to the optimal solution of~\eqref{eq:equMPC}.
This is the so-called \emph{receding horizon policy} that characterizes predictive control laws.

Obviously, the MPC controller~\eqref{eq:equMPC} will only be able to achieve the control objective of steering the system state $x(t)$ to the desired reference $(\xr, \ur)$ if the following two conditions are satisfied:
\begin{enumerate}[leftmargin=2.2em, label=(\textit{c.\roman*})]
    \item The initial state $x(0)$ belongs to the \emph{domain of attraction} of the MPC controller, that is, to the set of states for which the MPC control law asymptotically steers the system to $\xr$.
        For a well-posed MPC, the domain of attraction coincides with the set of states $x(t)$ for which optimization problem~\eqref{eq:equMPC} is feasible~\cite{RMD17}. \label{cond:domain}
    \item The desired reference $(\xr, \ur)$ is an admissible steady state of the system (see Definition~\ref{def:ass}). This condition might be limiting in a practical setting, where various reasons may lead to the selection of a non-admissible reference, e.g., a user with insufficient knowledge of the system. \label{cond:ref}
\end{enumerate}

Therefore, in a practical setting, we are interested in reducing the impact of the above conditions \ref{cond:domain} and \ref{cond:ref}.
In this regard, the MPC formulation~\eqref{eq:equMPC} is not an ideal choice.
Indeed, the main issue with~\eqref{eq:equMPC} is that its feasibility region tends to be small.
Note that, due to \eqref{eq:equMPC:terminal}, only states $x(t)$ that can reach the reference $\xr$ in $N$ steps may belong to the feasibility region of~\eqref{eq:equMPC}.
If the prediction horizon is chosen too small or the constraints are too restrictive, then it is very likely that the feasibility region (and thus the domain of attraction of the controller) is too small for its practical use.
This issue becomes even more apparent when considering the possibility of online changes of the reference, since a change of the reference may lead to a loss of feasibility of the MPC controller if it is too far away from the current state of the system.
Instead, we are interested in an MPC formulation whose domain of attraction is as large as possible, to mitigate \ref{cond:domain}, and that can inherently deal with non-attainable references, to deal with \ref{cond:ref}.

\section{MPC for tracking piecewise-constant references} \label{sec:MPCT}

An ideal choice to mitigate the issues raised in the previous section is the MPC formulation known as \emph{MPC for Tracking} (MPCT)~\cite{Limon_A_2008, Ferramosca_A_2009}.
MPCT has several variations, including non-linear~\cite{Limon_TAC_2018}, periodic~\cite{Limon_MPCTP_2016,Limon_IFAC_2012}, economic~\cite{Limon_JPC_2014} or robust~\cite{Limon_JPC_2010,Pereira_TAC_2016}.
In this section we focus on a variation of linear MPCT that considers a terminal equality constraint.
The reason for this choice is that this formulation is easier to pose, design and understand, providing us an ideal scenario to explain the main ideas and concepts behind the \emph{tracking} MPC formulations.
In the following sections we will take a look at two more recent variations of MPCT: periodic MPCT and \emph{harmonic} MPCT.

The idea of MPCT is to add an \emph{artificial reference} $(\xs, \us)$ as new decision variables of the MPC optimization problem, resulting in
\begin{subequations} \label{eq:MPCT} % MPCT
\begin{align}  
    \min\limits_{\substack{\vv{x}, \vv{u},\\ \xs, \us}} &\; \Sum{k = 0}{N-1} \left( \| x(k|t) - \xs \|^2_Q + \| u(k|t) - \us \|^2_R \right) + \Vo(\xs, \us; \xr, \ur) \\
    \st & \; x(0|t) = x(t), \label{eq:MPCT:initial} \\
        & \; x(k+1|t) = A x(k|t) + B u(k|t), \; k\in\N_0^{N-1}, \label{eq:MPCT:prediction} \\
        & \; \yLB \leq E x(k|t) + F u(k|t) \leq \yUB, \; k\in\N_0^{N-1}, \label{eq:MPCT:constraints} \\
        & \; x(N|t) = \xs, \label{eq:MPCT:terminal} \\
        & \; \xs = A \xs + B \us, \label{eq:MPCT:art_ref:steady_state}\\
        & \; \yLB + \sigma \ones{\ny} \leq E \xs + F \us \leq \yUB - \sigma \ones{\ny}, \label{eq:MPCT:art_ref:constraints}
\end{align}
\end{subequations}
where $\sigma > 0$ and $\Vo(\cdot)$, known as the \emph{offset cost function}, is assumed to be a convex function such that $(\xr, \ur) = \arg\min_{\xs, \us} \Vo (\xs, \us; \xr, \ur)$.
The typical choice for the offset cost function is to take 
\begin{equation} \label{eq:Vo:quadratic}
\Vo(\xs, \us; \xr, \ur) = \| \xs - \xr \|^2_T + \| \us - \ur \|^2_S,
\end{equation}
where $T \in \R^{\nx \times \nx}$ and $S \in \R^{\nu \times \nu}$ are positive definite.
The offset cost plays the part of a terminal ingredient between the artificial reference and the desired reference, whereas the summation of \emph{stage costs} $\| x(k|t) - \xs \|^2_Q + \| u(k|t) - \us \|^2_R$ along the prediction horizon $N$ are the classical terms of the MPC cost function, cf.~\eqref{eq:equMPC:cost_function}.
Constraint \eqref{eq:MPCT:terminal} is imposing the predicted terminal state $x(N|t)$ to be equal to the artificial reference $\xs$, which by constraints \eqref{eq:MPCT:art_ref:steady_state} and \eqref{eq:MPCT:art_ref:constraints} is required to be an admissible steady state of the system.
The positive scalar $\sigma$ is taken arbitrarily small and is included to avoid a possible loss of controllability in the event in which there are active constraints at an equilibrium point \cite{Limon_A_2008}.

\begin{figure}[t]
    \centering
    \includegraphics[width=0.65\linewidth]{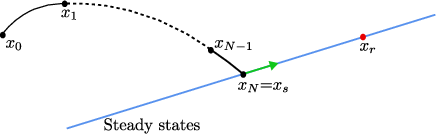}
\caption{Intuitive representation of the MPCT artificial reference. The blue line represents the space of steady states of the system and the red dot the desired reference~$\xr$.}
    \label{fig:MPCT:art_ref}
\end{figure}

To reiterate, the idea behind MPCT is to introduce a \emph{proxy} of the reference in the form of the decision variables $(\xs, \us)$.
The cost function penalizes, on one hand, the discrepancy between the predicted states and the artificial reference by means of the stage costs, and on the other, the distance between the artificial reference and the desired reference by means of the offset cost.
The idea is that the artificial reference $(\xs, \us)$ will converge towards the desired reference $(\xr, \ur)$, and that the predicted states will, in turn, converge towards the artificial reference.
This idea is illustrated in Figure~\ref{fig:MPCT:art_ref}.

A first clear advantage of MPCT \eqref{eq:MPCT}, when compared to \eqref{eq:equMPC}, is that its feasibility region can be much larger.
Intuitively, this is because in the MPCT formulation~\eqref{eq:MPCT}, the current state $x(t)$ only needs to be able to reach \emph{any} admissible steady state of the system in $N$ steps, as illustrated in Figure~\ref{fig:MPCT:art_ref}.
In contrast, problem~\eqref{eq:equMPC} is unfeasible if the current state $x(t)$ cannot reach the \emph{desired reference} in $N$ steps.
As a result, MPCT~\eqref{eq:MPCT} typically has a much larger domain of attraction than~\eqref{eq:equMPC}, especially when using small prediction horizons\footnote{For large prediction horizons the feasibility region of both formulations converge to the same maximal stabilizable invariant set.}.

Another useful feature of the MPCT formulation~\eqref{eq:MPCT} is that it retains recursive feasibility, under nominal conditions, even in the event of a change of the reference $(\xr, \ur)$ between sample times.
That is, if optimization problem~\eqref{eq:MPCT} is feasible for the current state $x(t)$, then the MPCT controller remains feasible for all future sample times.
In fact, recursive feasibility is maintained even if the reference is not an admissible steady state of the system; a useful feature in a practical setting, where the desired reference may be inadmissible, for instance, due to it being set by a higher-level part of the control architecture.
The following theorem formalizes the recursive feasibility of the MPCT formulation~\eqref{eq:MPCT}.
Its proof can be found in \cite{Limon_A_2008, Limon_TAC_2018}.

\begin{theorem}[Recursive feasibility of MPCT] \label{theo:MPCT:feasibility}
    Let~$x(t)$ belong to the feasibility region of the MPCT formulation~\eqref{eq:MPCT}. 
    Let $\tilde{\vv{x}}$, $\tilde{\vv{u}}$, $\tilde{x}_s$ and $\tilde{u}_s$ be any feasible solution of~\eqref{eq:MPCT} for a given reference $(\xr(t), \ur(t))$.
    Then, the successor state $A x(t) + B \tilde{u}(0|t)$ belongs to the feasibility region of~\eqref{eq:MPCT} for any reference $(\xr(t+1), \ur(t+1))$.
\end{theorem}

As in standard MPC, the MPCT controller steers the system to the fixed reference $(\xr, \ur)$ if the reference is an admissible steady state of the system.
However, an interesting property of MPCT is that if the reference is non-admissible (i.e., it is not a steady state of the system or it does not satisfy the system constraints), then the closed-loop system asymptotically converges to the \emph{optimal reachable reference} $(\xo, \uo)$, which we formally define in the following definition.

\begin{definition}[Optimal reachable reference] \label{def:MPCT:orr}
Given a reference $(\xr, \ur) \in \R^\nx \times \R^\nu$, we define the \emph{optimal reachable reference} of the MPCT formulation~\eqref{eq:MPCT} as the unique solution $(\xo, \uo) \in \R^\nx \times \R^\nu$ of the strongly convex optimization problem
\begin{subequations} \label{eq:MPCT:orr}
\begin{align}
    (\xo, \uo) = \arg\min\limits_{x, u} &\; \Vo(x, u; \xr, \ur) \\
                               \st & \; x = A x + B u, \label{eq:MPCT:orr:steady_state}\\
                                   & \; \yLB + \sigma \ones{\ny} \leq E x + F u \leq \yUB - \sigma \ones{\ny}. \label{eq:MPCT:orr:constraints}
\end{align}
\end{subequations}
\end{definition}

Note that if the reference $(\xr, \ur)$ is a strictly admissible steady state, then there exists a sufficiently small $\sigma > 0$ such that the optimal reachable reference satisfies $(\xo, \uo) = (\xr, \ur)$.
In this case, the closed-loop system asymptotically converges to the desired reference.
Otherwise, it converges to the closest strictly admissible steady state to the desired reference (\emph{strictly} due to $\sigma > 0$), as measured by the offset cost function $\Vo(\cdot)$.
The following theorems formalizes the asymptotic stability of the MPCT formulation~\eqref{eq:MPCT}.
Its proof can be found in \cite{Limon_A_2008, Limon_TAC_2018}.

\begin{theorem}[Asymptotic stability of MPCT] \label{theo:MPCT:stability}
    Let $(\xr, \ur)$ be a fixed desired reference.
    Assume that the initial state $x(0)$ belongs to the feasibility region of~\eqref{eq:MPCT} and that the prediction horizon $N$ is greater or equal to the controllability index of system~\eqref{eq:model}.
    Then, system~\eqref{eq:model} controlled with the control law of the MPCT formulation~\eqref{eq:MPCT} is stable, fulfills the system constraints for all $t \in \N$, and asymptotically converges to the optimal reachable reference $(\xo, \uo)$ given by Definition~\ref{def:MPCT:orr}.
\end{theorem}

\begin{remark}
    Note that Theorem~\ref{theo:MPCT:stability} requires the prediction horizon $N$ to be larger or equal to the controllability index of system~\eqref{eq:model}, i.e., the smallest  $j \in \N_1^{\nx}$ for which $\begin{bmatrix} B, A B, A^2 B, \dots, A^{j-1} B \end{bmatrix}$ has rank $\nx$.
    In practice, this often means that $N \geq \nx$.
\end{remark}

As in standard MPC, the optimization problem of the MPCT formulation \eqref{eq:MPCT}, when using a quadratic offset cost function \eqref{eq:Vo:quadratic}, is also a simple QP problem that can be solved online using one of the many available efficient QP solvers \cite{Stellato_OSQP, ODonoghue_SCS_21}.
However, the inclusion of the artificial reference as decision variables in problem~\eqref{eq:MPCT} complicates the resulting QP problem when compared to standard MPC formulation such as~\eqref{eq:equMPC}, both in terms of the number of decision variables and of the sparsity pattern of the QP matrices.
Even so, problem~\eqref{eq:MPCT} is still a sparse QP problem that can be efficiently solved using any of the many available QP solvers from the literature, such as~\cite{Stellato_OSQP, ODonoghue_SCS_21}.
Additionally, sparse solvers that exploit the particular structures of the MPCT formulation~\eqref{eq:MPCT} have recently been proposed in~\cite{Krupa_TCST_2021, Gracia_ECC_effMPCT_24}; the latter based on the ADMM algorithm~\cite{Boyd_FTML_2011} and the former on the \emph{extended} ADMM algorithm~\cite{Cai_EADMM_2017}.
Both solvers are available in the Spcies toolbox for MATLAB~\cite{Spcies}.
The results from~\cite{Krupa_TCST_2021} indicate that the MPCT formulation~\eqref{eq:MPCT} can be solved in similar computation times to standard MPC formulations such as~\eqref{eq:equMPC}.

\vspace*{-0.8em}

\subsection{Case study: ball and plate system} \label{sec:MPCT:BaP}

In this chapter, we use the ball and plate system from~\cite[\S 5.7.2]{Krupa_Thesis_21} for the closed-loop results using the different tracking MPC formulations discussed in each section.

\begin{figure}[t]
    \centering
    \includegraphics[width=0.52\columnwidth]{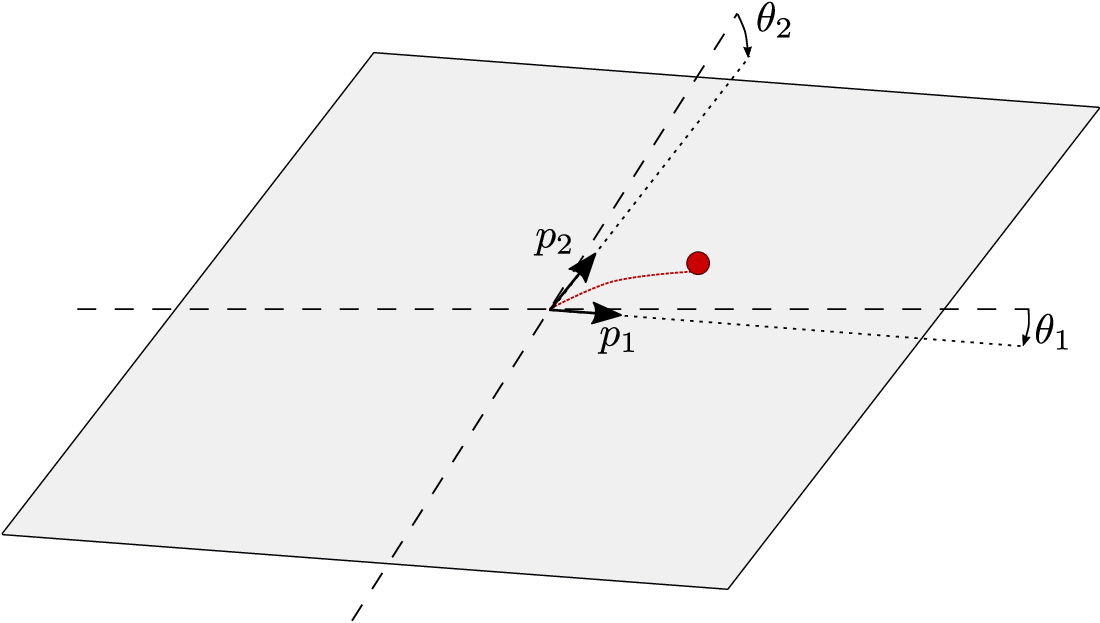}
    \caption{Ball and plate system.}
    \label{fig:BaP}
\end{figure}

The ball and plate system, shown in Figure~\ref{fig:BaP}, consists of a plate that is nominally in a horizontal position but whose inclination can be adjusted by applying an angular acceleration to either of its two main axes.
The control objective is to control the position of a solid ball that moves on the surface of the plate as a consequence of its inclination.
We assume that the ball is always in contact with the plate and that it does not slip when moving over its surface.
Under this assumption, the non-linear equations of the system are given by~\cite{Wang_ISA_2014}
\begin{subequations} \label{eq:BaP:nonlinear}
\begin{align}
    \ddot{\pBP}_1 &= \frac{m}{m + I_b/r^2} \left( \pBP_1 \dot{\aBP}_1^2 + \pBP_2 \dot{\aBP}_1 \dot{\aBP}_2 + g \sin{\aBP_1} \right), \\
    \ddot{\pBP}_2 &= \frac{m}{m + I_b/r^2} \left( \pBP_2 \dot{\aBP}_2^2 + \pBP_1 \dot{\aBP}_1 \dot{\aBP}_2 + g \sin{\aBP_2} \right),
\end{align}
\end{subequations}
where $\pBP_1$ and $\pBP_2$ are the coordinates of the ball on each of the axes of the plate\footnote{The center of the plate is considered the origin.}, $\dot{\pBP}_1$, $\dot{\pBP}_2$, $\ddot{\pBP}_1$ and $\ddot{\pBP}_2$ the corresponding velocity and acceleration of the ball, $\aBP_1$, $\dot{\aBP}_1$, $\aBP_2$ and $\dot{\aBP}_2$ are the angle and angular velocity of the plate in each of its axes, $m = 0.05$Kg is the mass of the ball, $r = 0.01$m its radius, $I_b = (2/5) m r^2 = 2 \cdot 10^{-6}$Kg$\cdot$m$^{2}$ its mass moment of inertia, and $g = 9.81$m/s$^{2}$ is the gravitational constant.

The state of the system is given by $x = (\pBP_1, \dot{\pBP}_1, \aBP_1, \dot{\aBP}_1, \pBP_2, \dot{\pBP}_2, \aBP_2, \dot{\aBP}_2)$ and the control input by $u = (\ddot{\aBP}_1, \ddot{\aBP}_2)$, that is, by the angular acceleration of each of the axes of the plate.
The constraints of the system are given by
\begin{equation} \label{eq:BaP:constraints}
    |\pBP_i | \leq 0.3\, \text{m}, \;
    |\dot{\pBP}_i| \leq 0.1\,\text{m/s}, \;
    |\aBP_i| \leq \frac{\pi}{4}\,\text{rad}, \;
    |\ddot{\aBP}_i| \leq 0.1\,\text{rad/s}^2, \; i \in \N_1^2.
\end{equation}
No constraint is imposed on $\dot{\aBP}_i$, $i \in \N_1^2$.
The box constraints~\eqref{eq:BaP:constraints} can be easily written as coupled input-state constraints~\eqref{eq:model:constraints}.
A linear model~\eqref{eq:model} is obtained by linearizing~\eqref{eq:BaP:nonlinear} around the operating point $(x, u) = (0, 0) \in \R^8 \times \R^2$, assuming a zero-order hold on the control inputs and taking a sample time of $0.2$ seconds.

\begin{figure}[t]
    \centering
    \begin{subfigure}[ht]{0.49\textwidth}
        \includegraphics[width=\linewidth]{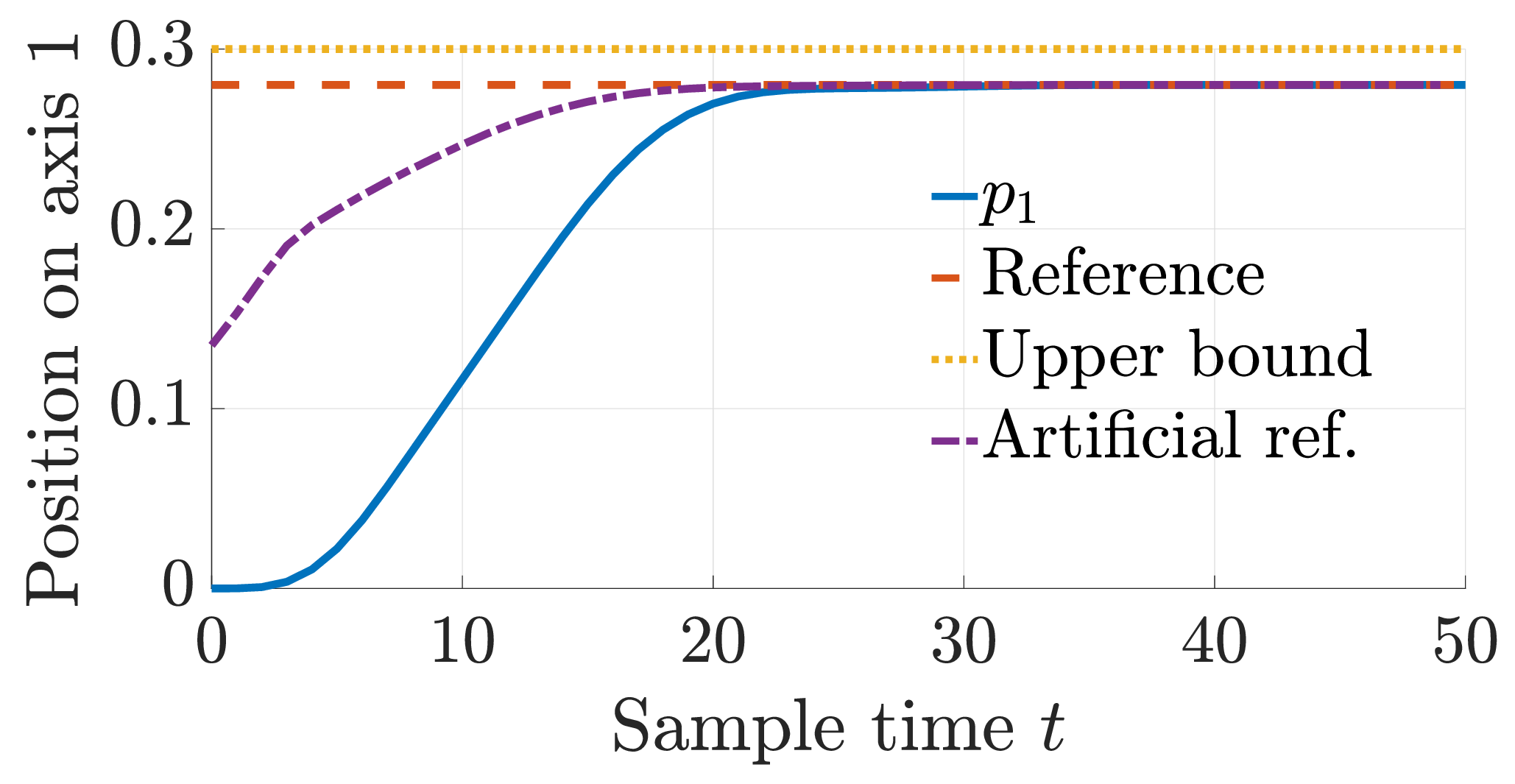}
        \caption{Trajectory for admissible reference.}
        \label{fig:BaP:MPCT:ad:traj}
    \end{subfigure}%
    \hfill%%
    \begin{subfigure}[ht]{0.49\textwidth}
        \includegraphics[width=\linewidth]{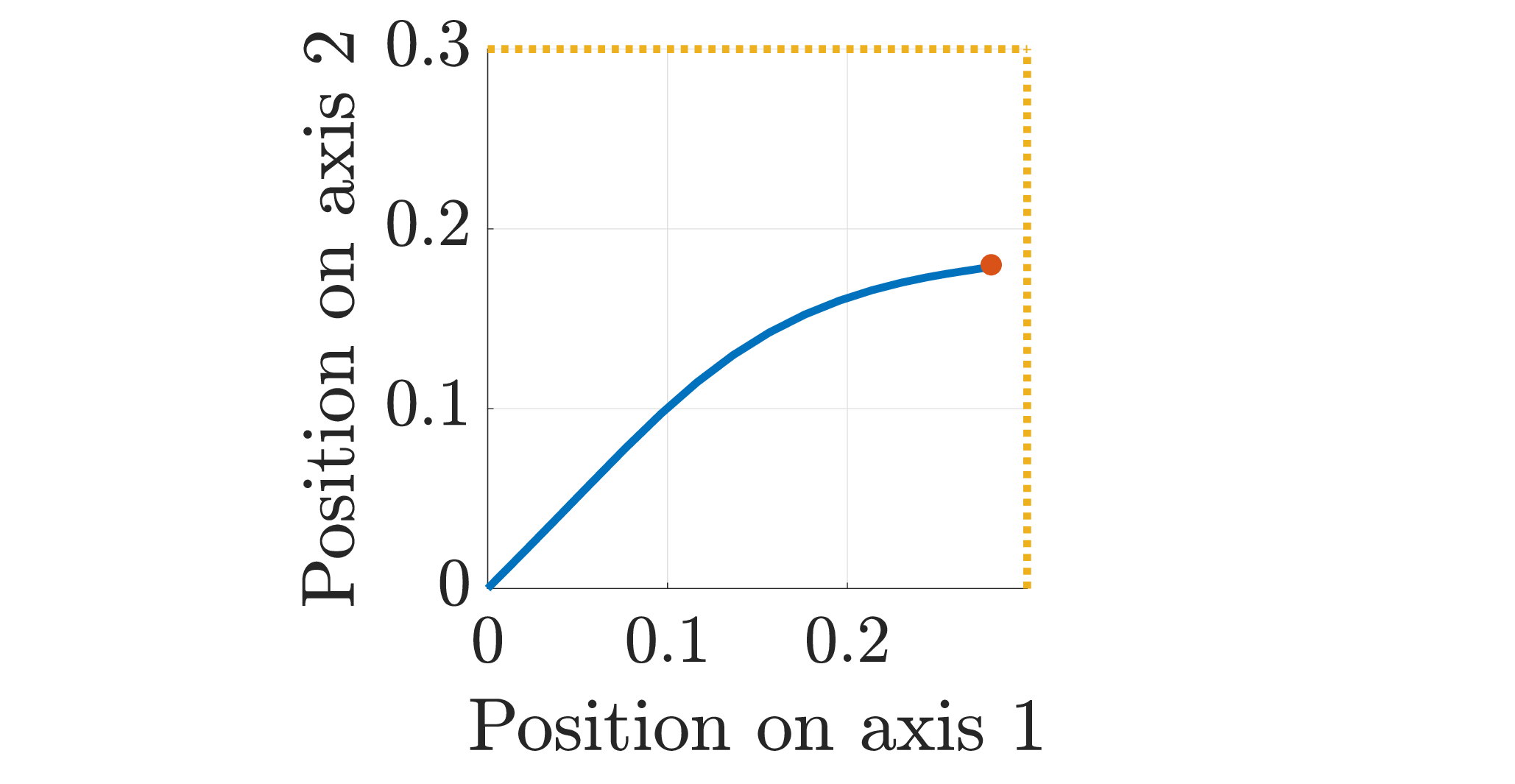}
        \caption{Position of ball for admissible reference.}
        \label{fig:BaP:MPCT:ad:pos}
    \end{subfigure}%

    \begin{subfigure}[ht]{0.49\textwidth}
        \includegraphics[width=\linewidth]{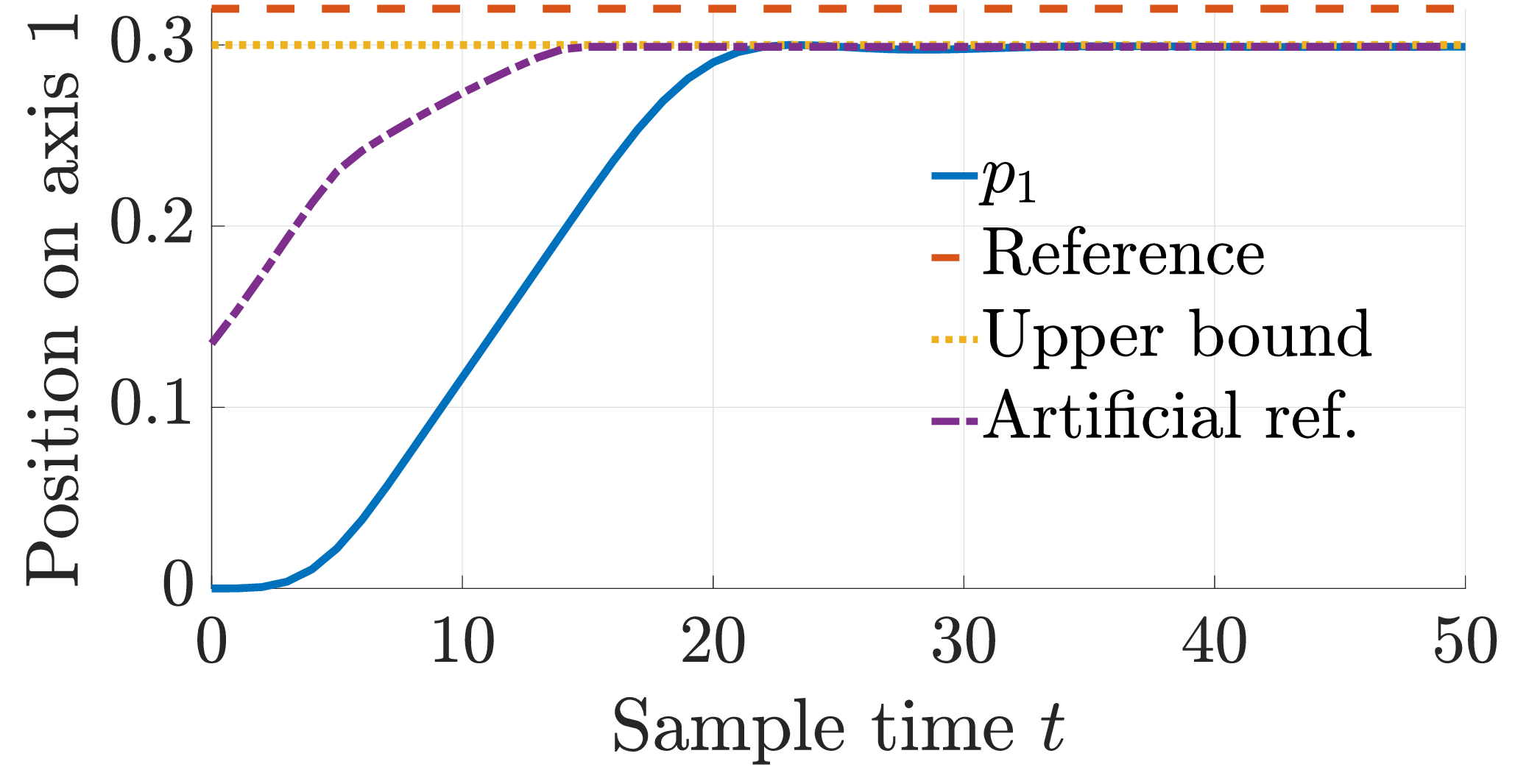}
        \caption{Trajectory for non-admissible reference.}
        \label{fig:BaP:MPCT:nonad:traj}
    \end{subfigure}%
    \hfill%%
    \begin{subfigure}[ht]{0.49\textwidth}
        \includegraphics[width=\linewidth]{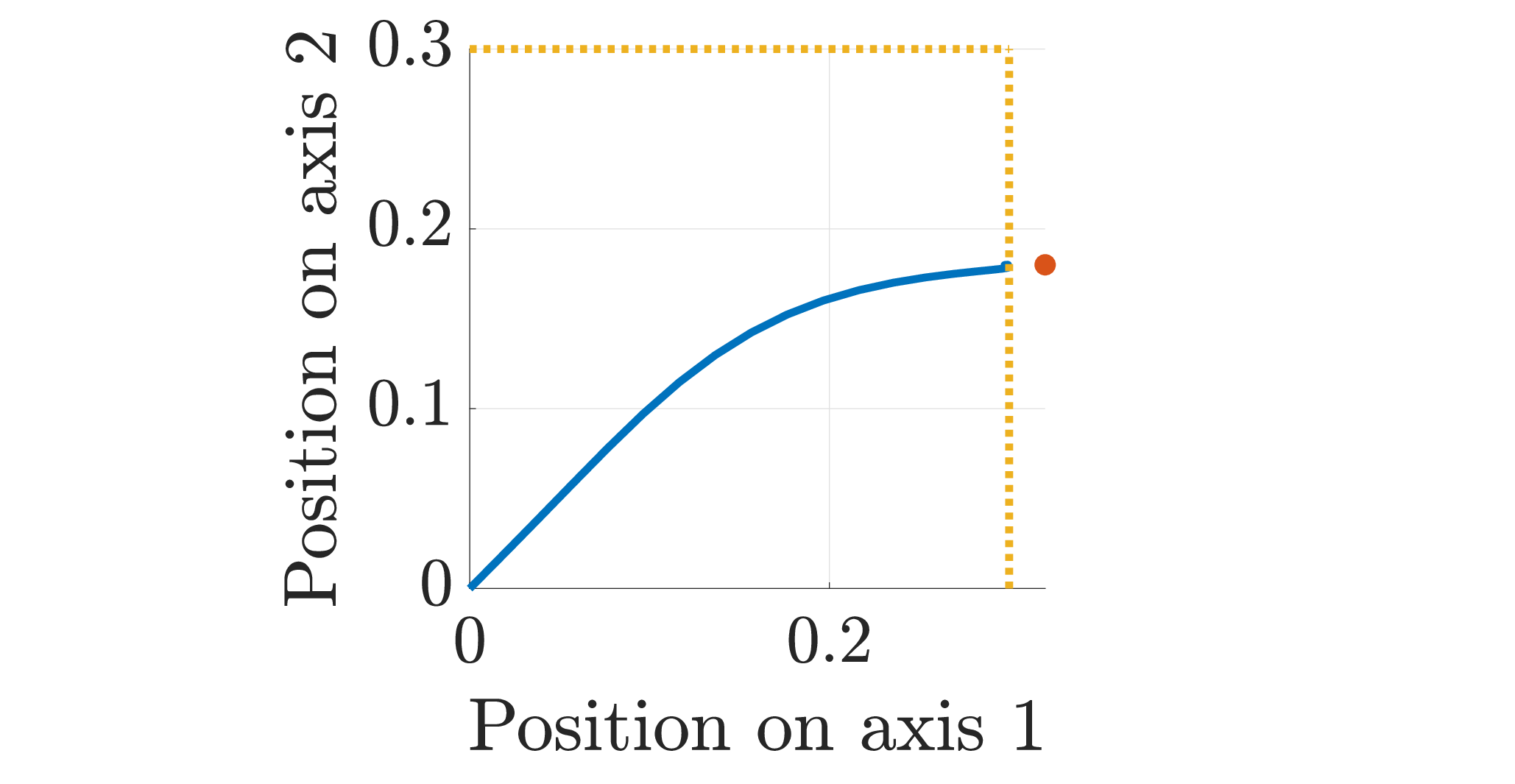}
        \caption{Position of ball for non-admissible reference.}
        \label{fig:BaP:MPCT:nonad:pos}
    \end{subfigure}%

    \caption{Closed-loop results of the ball and plate system using MPCT~\eqref{eq:MPCT}.}
    \label{fig:BaP:MPCT}
\end{figure}

Figure~\ref{fig:BaP:MPCT} shows the closed-loop results of the ball and plate system controlled using MPCT~\eqref{eq:MPCT} taking $N = 15$, 
$Q = \diag(10, 0.05, 0.05, 0.05, 10, 0.05, 0.05, 0.05)$, $R = \diag(0.5, 0.5)$, $T = N Q$ and $S = N R$.
Figures~\ref{fig:BaP:MPCT:ad:traj} and~\ref{fig:BaP:MPCT:ad:pos} show the trajectory of the position of the ball over time and the trajectory of the ball on the plate, respectively, for an admissible desired reference.
Figures~\ref{fig:BaP:MPCT:nonad:traj} and~\ref{fig:BaP:MPCT:nonad:pos} show the same results, but for a non-admissible reference.
Figures~\ref{fig:BaP:MPCT:ad:traj} and~\ref{fig:BaP:MPCT:nonad:traj} include the trajectory of the artificial reference for the position $p_1$ of the ball.
The results show how the MPCT formulation~\eqref{eq:MPCT} steers the system to the optimal reachable reference when the desired reference in non-admissible.
In this case, since $T$ and $S$ are diagonal, the optimal reachable reference is the admissible steady state with the smallest weighted Euclidean distance to the desired reference.

\begin{remark} \label{rem:results:Spcies}
This chapter does not focus on the computational aspects of solving the tracking MPC formulations.
However, for completeness, we note that the results shown in Figure~\ref{fig:BaP:MPCT} are obtained using the solvers available in version \texttt{v0.3.11} of the Spcies toolbox for MATLAB~\cite{Spcies}.
In particular, the MPCT formulation was solved using the ADMM solver~\cite{Gracia_ECC_effMPCT_24} for MPCT from the toolbox.
\end{remark}

\section{MPC for tracking periodic references} \label{sec:MPCT:periodic}

This section presents an extension of MPCT for tracking periodic references~\cite{Limon_MPCTP_2016,Limon_IFAC_2012}.
That is, we now consider the control objective of following a desired \emph{periodic} reference trajectory $(\xr(t), \ur(t))$ with period $\Tp \in \N$, i.e., a trajectory satisfying $\xr(t+\Tp) = \xr(t)$ and $\ur(t+\Tp) = \ur(t)$ for all $t \in \N$.
As in Definition~\ref{def:ass}, we define the notion of admissibility, but in this case for a trajectory.

\begin{definition}[Admissible trajectory] \label{def:ass:trajectory}
    A trajectory $(x(t), u(t))$, with $x(t) \in \R^\nx$ and $u(t) \in \R^\nu$, is said to be \emph{admissible} for system~\eqref{eq:model} subject to~\eqref{eq:model:constraints} if it satisfies $x(t+1) = A x(t) + B u(t)$ and $\yLB \leq E x(t) + F u(t) \leq \yUB$ for all $t \in \N$. 
    Furthermore, we say that it is \emph{strictly admissible} if it satisfies $\yLB < E x(t) + F u(t) < \yUB$ for all $t \in \N$.
\end{definition}

As in the case of the MPCT formulation~\eqref{eq:MPCT} for tracking piecewise-constant references, periodic MPCT can be seen as an extension of the classical periodic MPC formulations, which are often related to economic MPC~\cite{Zanon_CDC_2013, Zanon_IFAC_2017, Muller_AUT_2016}.
These formulations can also suffer from the same issues discussed in Section~\ref{sec:intro}, i.e., insuficiently large domain of attraction and loss of feasibility if the periodic reference is changed online.
That is, the periodic reference at time $t+1$ is not the shifted reference of the previous sample time, but some new periodic reference instead.
The periodic MPCT formulation mitigates both of these issues using the same idea as MPCT~\eqref{eq:MPCT}: by introducing an artificial reference.
In this case, however, instead of a steady state artificial reference, the periodic MPCT formulation uses a periodic artificial reference whose period coincides with the period of the desired reference.
That is, to add a periodic admissible reference trajectory $(\xs(k|t), \us(k|t))$ as decision variable of the optimization problem, resulting in:
\begin{subequations} \label{eq:perMPCT}
\begin{align}  
    \min\limits_{\substack{\vv{x}, \vv{u},\\ \vv{\xs}, \vv{\us}}} &\; \Sum{k = 0}{N-1} \left( \| x(k|t) {-} \xs(k|t) \|^2_Q {+} \| u(k|t) {-} \us(k|t) \|^2_R \right) + \Vp(\vv{\xs}, \vv{\us}; \xr(t), \ur(t)) \\
    \st & \; x(0|t) = x(t), \label{eq:perMPCT:initial} \\
        & \; x(k+1|t) = A x(k|t) + B u(k|t), \; k\in\N_0^{N-1}, \label{eq:perMPCT:prediction} \\
        & \; \yLB \leq E x(k|t) + F u(k|t) \leq \yUB, \; k\in\N_0^{N-1}, \label{eq:perMPCT:constraints} \\
        & \; x(N|t) = \xs(N|t), \label{eq:perMPCT:terminal} \\
        & \; \xs(0|t) = A \xs(\Tp-1|t) + B \us(\Tp-1|t), \label{eq:perMPCT:art_ref:periodic}\\
        & \; \xs(k+1|t) = A \xs(k|t) + B \us(k|t), \; k\in\N_0^{\Tp-2}, \label{eq:perMPCT:art_ref:prediction}\\
        & \; \yLB + \sigma \ones{\ny} \leq E \xs(k|t) + F \us(k|t) \leq \yUB - \sigma \ones{\ny}, \; k\in\N_0^{\Tp-1}, \label{eq:perMPCT:art_ref:constraints}
\end{align}
\end{subequations}
where $\vv{\xs} \doteq (\xs(0|t), \dots \xs(\Tp-1|t))$ and $\vv{\us} \doteq (\us(0|t), \dots \us(\Tp-1|t))$, with $\xs(k|t) \in \R^\nx$ and $\us(k|t) \in \R^\nu$ for all $k \in \N_0^{\Tp-1}$.
Constraint~\eqref{eq:perMPCT:art_ref:periodic} forces the artificial reference to be a periodic trajectory with period $\Tp$.
The offset cost function $\Vp(\cdot)$ penalizes the distance between the periodic artificial reference and the desired periodic reference along the length of their periods; the typical choice is
\begin{equation*}
   \Vp(\vv{\xs}, \vv{\us}; \xr(t), \ur(t)) = \Sum{k=0}{\Tp-1} \| \xs(k|t) - \xr(k+t) \|^2_T + \| \us(k|t) - \ur(k+t) \|^2_S,
\end{equation*}
where $T \in \R^{\nx \times \nx}$ and $S \in \R^{\nu \times \nu}$ are positive definite.
The stage cost term of the cost function and the other constraints of~\eqref{eq:perMPCT} resemble the ones from the MPCT~\eqref{eq:MPCT}.
Indeed, both formulations force the terminal state $x(N|t)$ to reach the artificial reference, cf.~\eqref{eq:MPCT:terminal} and~\eqref{eq:perMPCT:terminal}, and both impose the artificial reference to be admissible, cf.~\eqref{eq:MPCT:art_ref:steady_state}, \eqref{eq:MPCT:art_ref:constraints}, \eqref{eq:perMPCT:art_ref:prediction} and~\eqref{eq:perMPCT:art_ref:constraints}.

The periodic MPCT~\eqref{eq:perMPCT} shares the same two main properties of MPCT~\eqref{eq:MPCT}, namely, recursive feasibility and asymptotic stability.
The following theorems formally state these two properties.
Proofs can be found in~\cite{Limon_MPCTP_2016}.

\begin{theorem}[Recursive feasibility of periodic MPCT] \label{theo:perMPCT:feasibility}
    Let $x(t)$ belong to the feasibility region of the periodic MPCT formulation~\eqref{eq:perMPCT}. 
    Let $\tilde{\vv{x}}$, $\tilde{\vv{u}}$, $\vv{\tilde{x}_s}$ and $\vv{\tilde{u}_s}$ be any feasible solution of~\eqref{eq:perMPCT} for a given periodic reference trajectory $(\xr(t), \ur(t))$ with period $\Tp$.
    Then, at time $t+1$, the successor state $x(t+1) = A x(t) + B \tilde{u}(0|t)$ belongs to the feasibility region of~\eqref{eq:perMPCT} for any periodic reference trajectory with period $\Tp$, not necessarily equal to $(\xr(t), \ur(t))$.
\end{theorem}

Note that the periodic reference can be changed between sample times, but not the period of the reference.
In the periodic case, the MPCT formulation asymptotically steers the system to the optimal reachable \emph{periodic} reference, which is a unique periodic trajectory determined by the offset cost function $\Vp(\cdot)$.

\begin{definition}[Optimal reachable periodic reference] \label{def:perMPCT:orr}
    Given a periodic reference trajectory $(\xr(t), \ur(t))$ with period $\Tp$, we define the \emph{optimal reachable periodic reference} of the MPCT formulation~\eqref{eq:perMPCT} as the unique solution $(\vv{\xo}, \vv{\uo})$ of the strongly convex optimization problem
\begin{subequations} \label{eq:perMPCT:orr}
\begin{align}
    (\vv{\xo}, \vv{\uo}) = \arg\min\limits_{\vv{x}, \vv{u}} &\; \Vp(\vv{x}, \vv{u}; \xr(0), \ur(0))  \\
    \st & \; x(0) = A x(\Tp-1) + B u(\Tp-1) \label{eq:perMPCT:orr:periodic}\\
        & \; x(k+1) = A x(k) + B u(k), \; k \in \N_0^{\Tp-2} \label{eq:perMPCT:orr:steady_state}\\
                                   & \; \yLB + \sigma \ones{\ny} \leq E x(t) + F u(t) \leq \yUB - \sigma \ones{\ny}, \; k \in \N_0^{\Tp-1}, \label{eq:perMPCT:orr:constraints}
\end{align}
\end{subequations}
where for $k \in \N_0^{\Tp-1}$, $\xo(k) \in \R^\nx$, $\uo(i) \in \R^\nu$, $\vv{\xo} \doteq (\xo(0), \xo(1), \dots \xo(\Tp-1))$, and $\vv{\uo} \doteq (\uo(0), \uo(1), \dots \uo(\Tp-1))$.
The optimal solution $(\vv{\xo}, \vv{\uo})$ defines a periodic trajectory $(\xo(t), \uo(t))$ with period $\Tp$.
\end{definition}

\begin{theorem}[Asymptotic stability of periodic MPCT] \label{theo:perMPCT:stability}
    Let $(\xr(t), \ur(t))$ be the desired periodic reference trajectory.
    Assume that the initial state $x(0)$ belongs to the feasibility region of~\eqref{eq:perMPCT} and that the prediction horizon $N$ is greater or equal to the controllability index of system~\eqref{eq:model}.
    Then, system~\eqref{eq:model} controlled with the control law of the periodic MPCT formulation~\eqref{eq:perMPCT} is stable, fulfills the system constraints for all $t \in \N$, and asymptotically converges to the optimal reachable periodic reference $(\xo(t), \uo(t))$ given by Definition~\ref{def:perMPCT:orr}.
\end{theorem}

\begin{remark}
    As with the MPCT formulation~\eqref{eq:MPCT}, the periodic MPCT formulation~\eqref{eq:perMPCT} will steer the system to the desired periodic reference trajectory if it is strictly admissible\footnote{For some sufficiently small $\sigma > 0$.} (see Definition~\ref{def:ass:trajectory}).
    Otherwise, it will steer the system to the ``closest'' admissible trajectory of the system, i.e., to the optimal reachable periodic reference $(\xo(t), \uo(t))$ given by Definition~\ref{def:perMPCT:orr}.
\end{remark}

The periodic MPCT formulation~\eqref{eq:perMPCT} is a QP problem, and can thus be solved using any of the many available QP solvers from the literature \cite{Stellato_OSQP, ODonoghue_SCS_21}.
However, we note that one of its drawbacks is that the number of decision variables and constraints of its optimization problem grows with the length of the period $\Tp$.
Indeed, constraints~\eqref{eq:perMPCT:art_ref:prediction} and~\eqref{eq:perMPCT:art_ref:constraints} impose the dynamics and constraints of the system on the artificial periodic reference throughout its period $\Tp$.
Note that a requirement of~\eqref{eq:perMPCT} is that the artificial periodic reference has the same period as the desired periodic reference.
Therefore, the complexity of the optimization problem depends on the period of the reference to be tracked.
This may lead to non-viable computation times and memory requirements for its online implementation if the prediction horizon is too large, especially when considering its implementation in embedded systems.

\subsection{Applying periodic MPCT to the ball and plate system} \label{sec:perMPCT:BaP}

We now present results applying the periodic MPCT formulation~\eqref{eq:perMPCT} to the ball and plate system presented in Section~\ref{sec:MPCT:BaP}.
As in Section~\ref{sec:MPCT:BaP}, we show closed-loop results for both an admissible and a non-admissible periodic reference trajectory, both with $\Tp = 25$.
The ingredients of the periodic MPCT formulation~\eqref{eq:perMPCT} are taken as $N = 15$ $Q = \diag(10, 0.05, 0.05, 0.05, 10, 0.05, 0.05, 0.05)$, $R = \diag(0.5, 0.5)$, $T = Q$ and $S = R$.
Note that we take the same $N$, $Q$ and $R$ used in Section~\ref{sec:MPCT:BaP} for the MPCT formulation~\eqref{eq:MPCT}.
We take the period of the artificial reference as $\Tp = 25$.

\begin{figure}[t]
    \centering
    \begin{subfigure}[ht]{0.49\textwidth}
        \includegraphics[width=\linewidth]{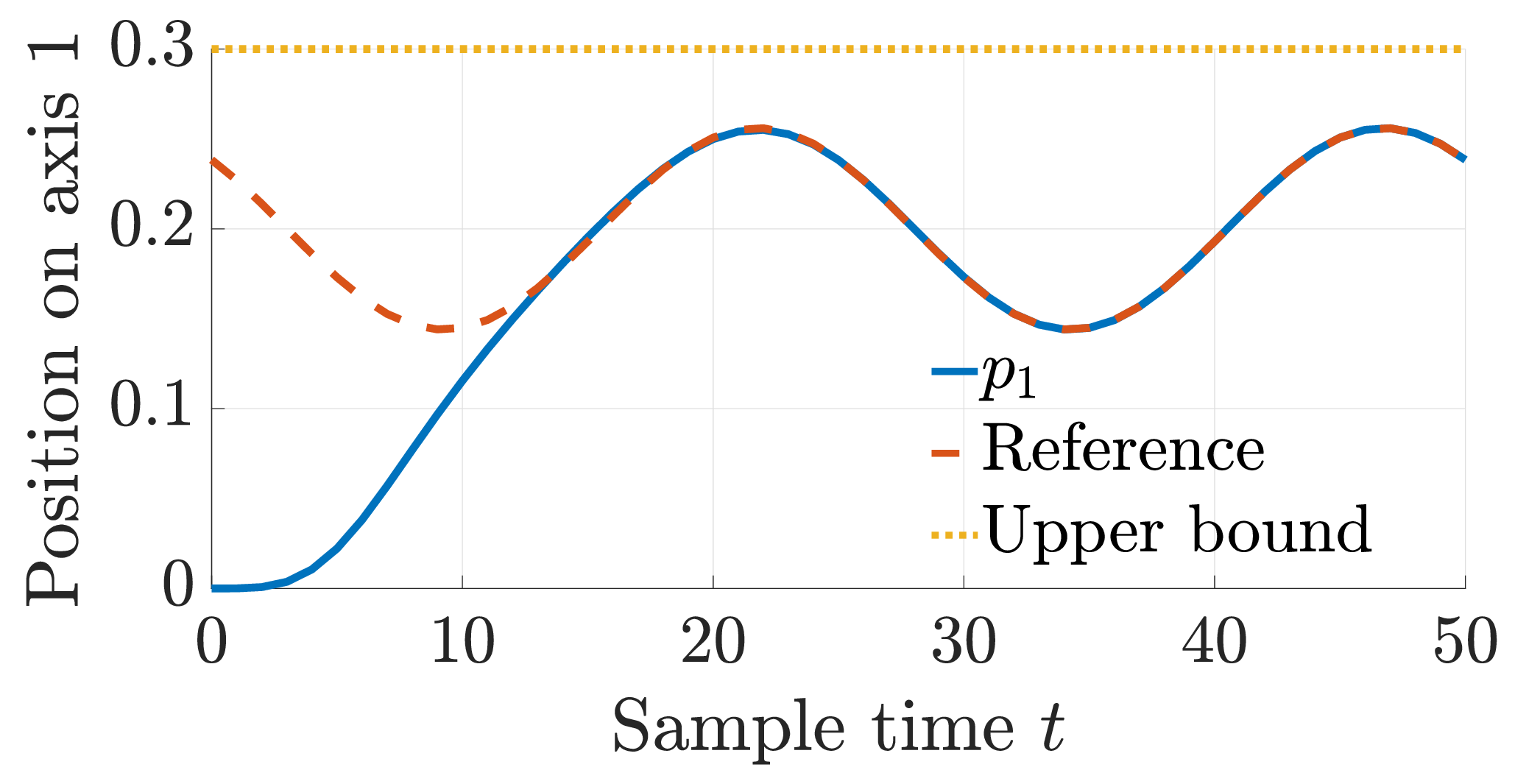}
        \caption{Trajectory for admissible reference.}
        \label{fig:BaP:perMPCT:ad:traj}
    \end{subfigure}%
    \hfill%%
    \begin{subfigure}[ht]{0.49\textwidth}
        \includegraphics[width=\linewidth]{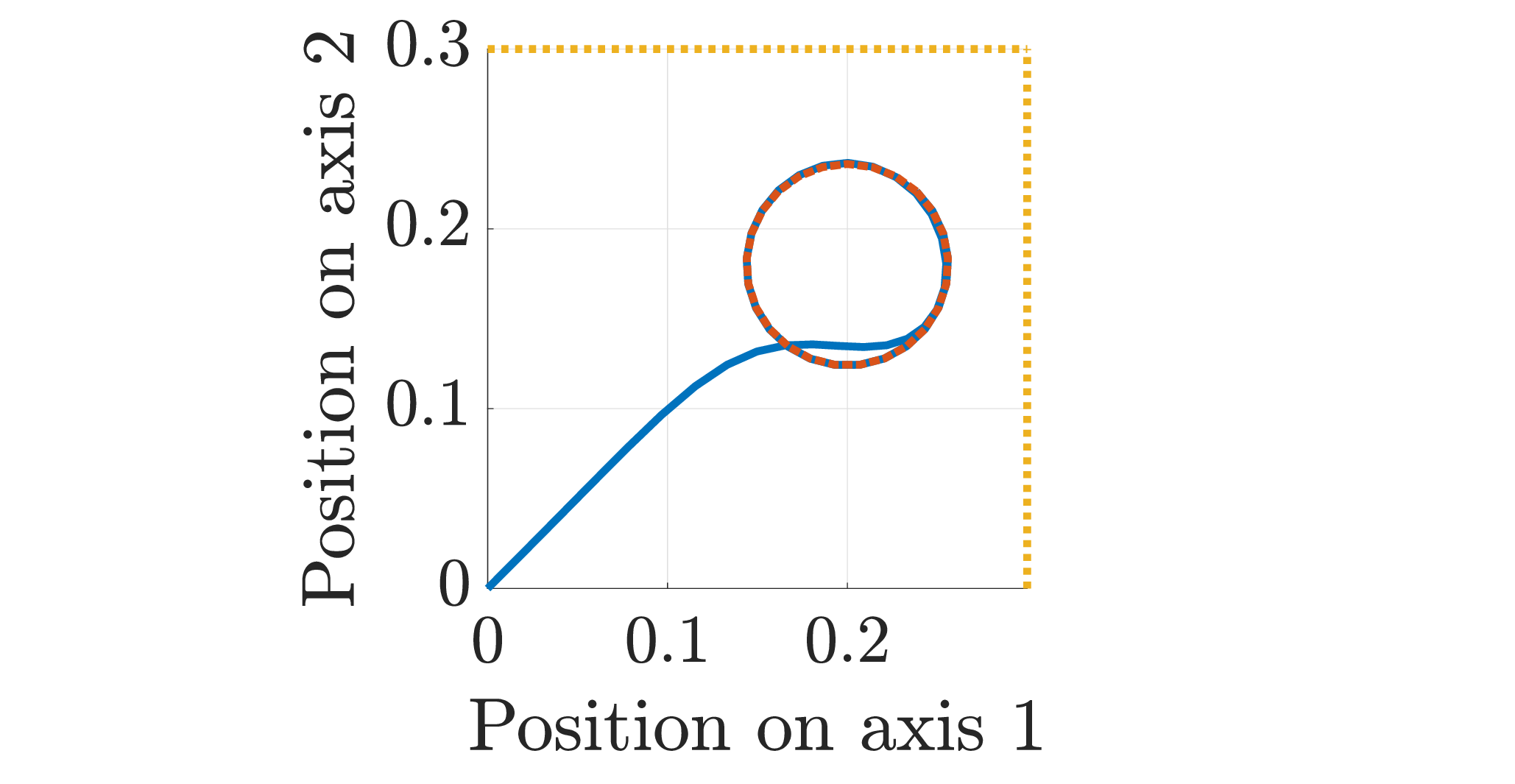}
        \caption{Position of ball for admissible reference.}
        \label{fig:BaP:perMPCT:ad:pos}
    \end{subfigure}%

    \begin{subfigure}[ht]{0.49\textwidth}
        \includegraphics[width=\linewidth]{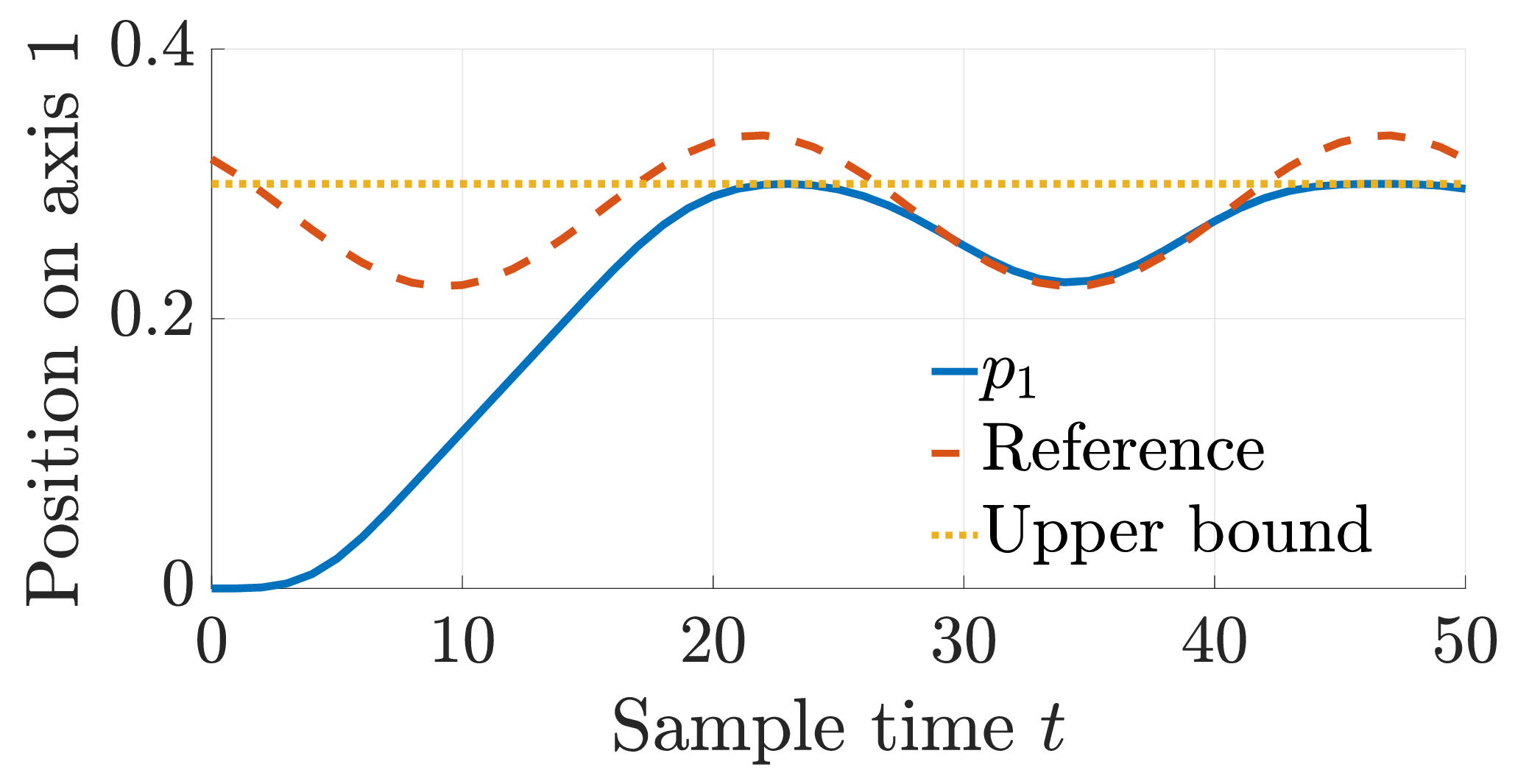}
        \caption{Trajectory for non-admissible reference.}
        \label{fig:BaP:perMPCT:nonad:traj}
    \end{subfigure}%
    \hfill%%
    \begin{subfigure}[ht]{0.49\textwidth}
        \includegraphics[width=\linewidth]{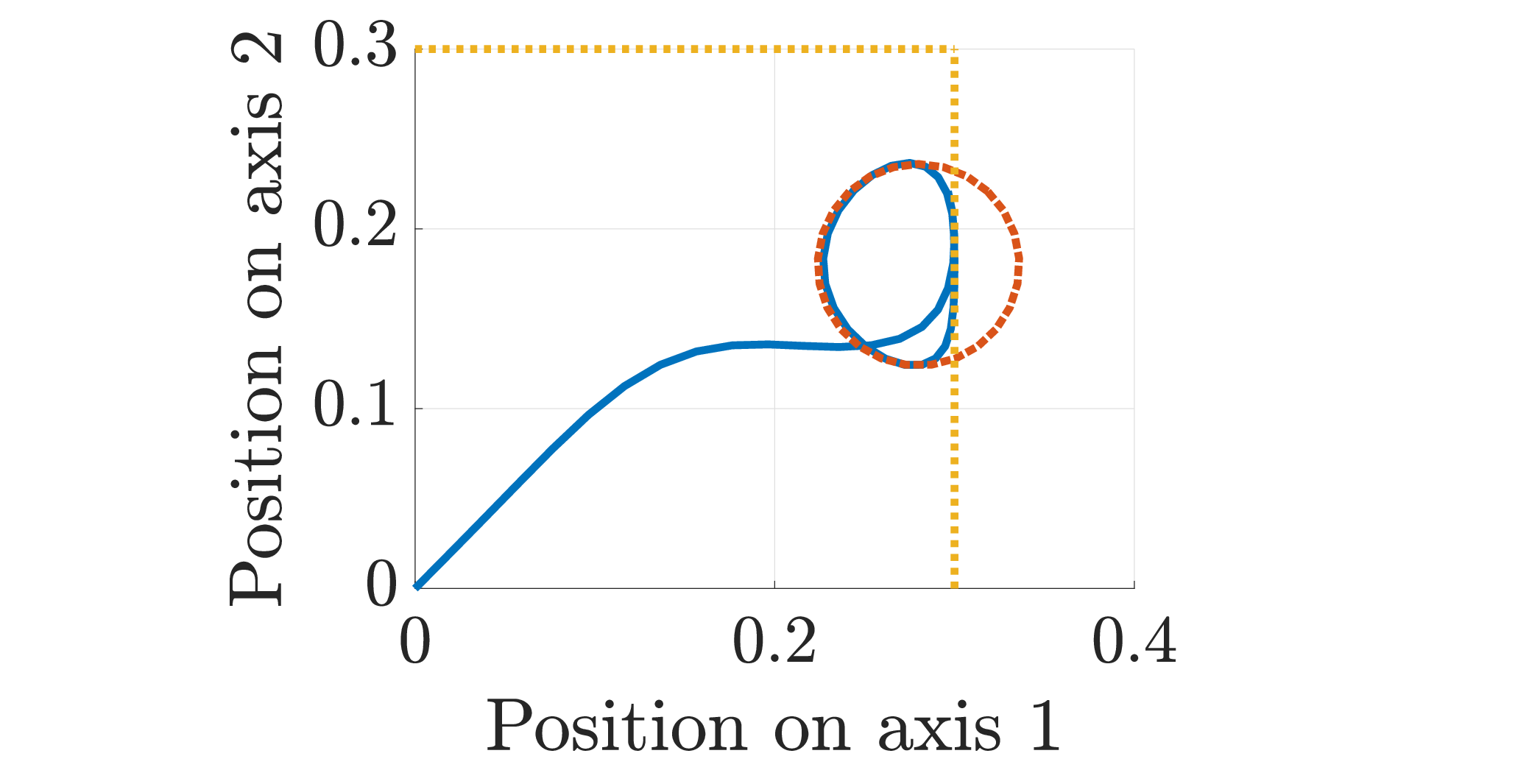}
        \caption{Position of ball for non-admissible reference.}
        \label{fig:BaP:perMPCT:nonad:pos}
    \end{subfigure}%

    \caption{Closed-loop results of the ball and plate system using periodic MPCT~\eqref{eq:perMPCT}.}
    \label{fig:BaP:perMPCT}
\end{figure}

Figure~\ref{fig:BaP:perMPCT} shows the closed-loop results.
Again, note that the periodic MPCT formulation~\eqref{eq:perMPCT} steers the system to the desired periodic reference when it is admissible.
Otherwise, it steers it to the optimal reachable periodic reference.
The non-admissible periodic reference used in Figures~\ref{fig:BaP:perMPCT:nonad:traj} and~\ref{fig:BaP:perMPCT:nonad:pos} is in fact only partially non-admissible.
Note how the periodic MPCT controller tracks the reference when it is admissible, and tries to be as close as possible to it when it violates the constraints.
Closeness in this case is also given by Euclidean distance, since matrices $T$ and $S$ are diagonal, although other measured of distance can be considered by taking non-diagonal positive definite matrices instead.

\begin{remark}
    For the closed-loop results shown in Figure~\ref{fig:BaP:perMPCT}, periodic MPCT~\eqref{eq:perMPCT} is solved using the MATLAB interface of version \texttt{v0.6.2} of the OSQP solver~\cite{Stellato_OSQP}.
\end{remark}

\section{Harmonic MPC for tracking} \label{sec:MPCT:harmonic}

This section presents a recent tracking MPC formulation that is inspired by both the MPCT~\eqref{eq:MPCT} and periodic MPCT~\eqref{eq:perMPCT} formulations. 
The idea of this formulation is to use a parametrized harmonic signal as the artificial reference, i.e., to use a particular type of periodic trajectory as the artificial reference, instead of the generic trajectory used in~\eqref{eq:perMPCT}.
However, its control objective is to track a steady state, as in MPCT~\eqref{eq:MPCT}, instead of a periodic reference.

This new formulation, originally presented in~\cite{Krupa_TAC_2022,Krupa_CDC_19} and which we call \emph{harmonic} MPC (HMPC) for obvious reasons, has advantages and disadvantages when compared to the classical MPCT formulation~\eqref{eq:MPCT} and its periodic extension~\eqref{eq:perMPCT}.
Let us start by taking a look at an example that highlights a drawback of MPCT~\eqref{eq:MPCT} that motivates the HMPC formulation.

\begin{figure}[t]
    \centering
    \begin{subfigure}[ht]{0.48\textwidth}
        \includegraphics[width=\linewidth]{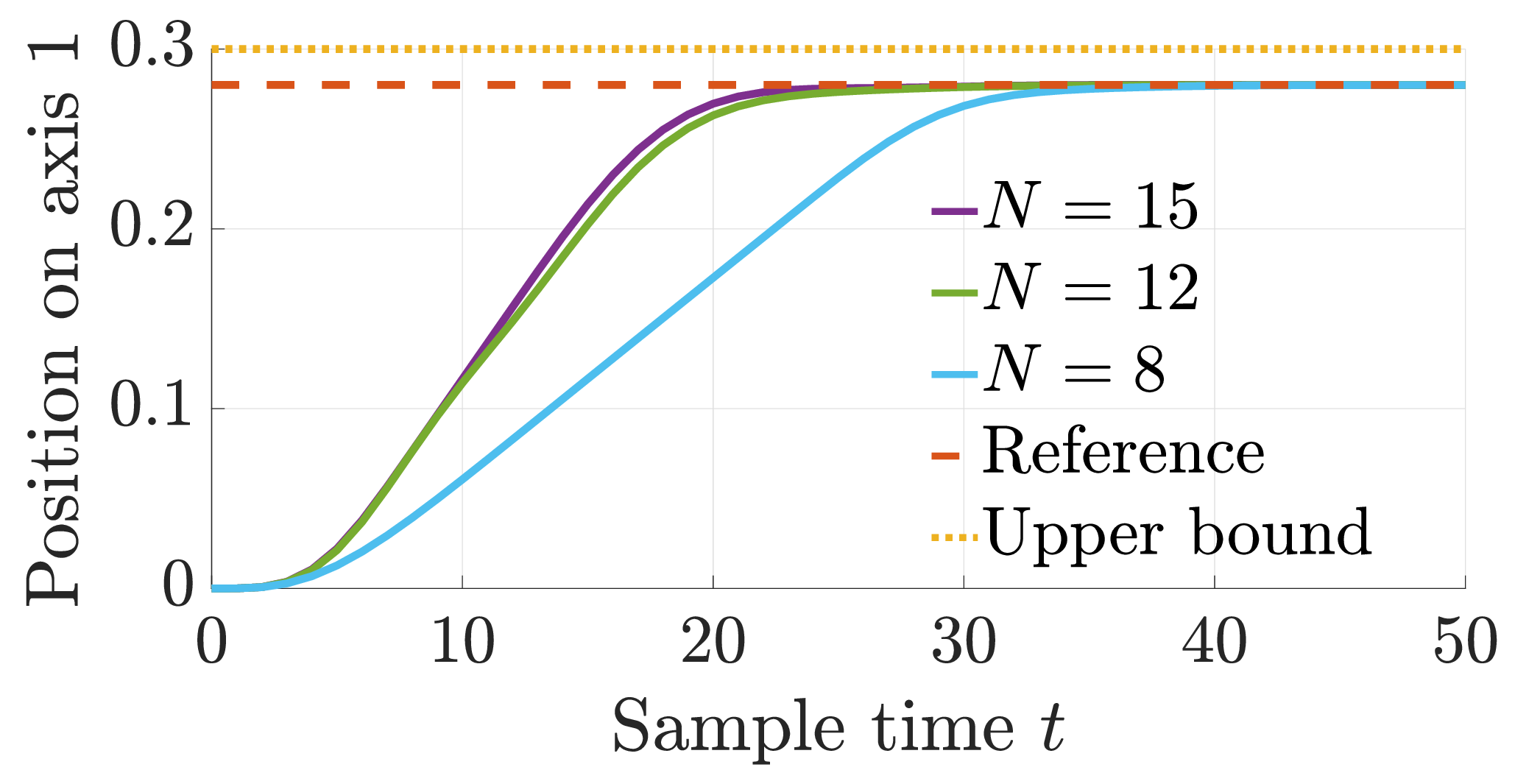}
        \caption{Trajectory of position of ball on axis 1 using MPCT~\eqref{eq:MPCT} with different prediction horizons.}
        \label{fig:HMPC:example:vel}
    \end{subfigure}%
    \hfill%%
    \begin{subfigure}[ht]{0.48\textwidth}
        \includegraphics[width=\linewidth]{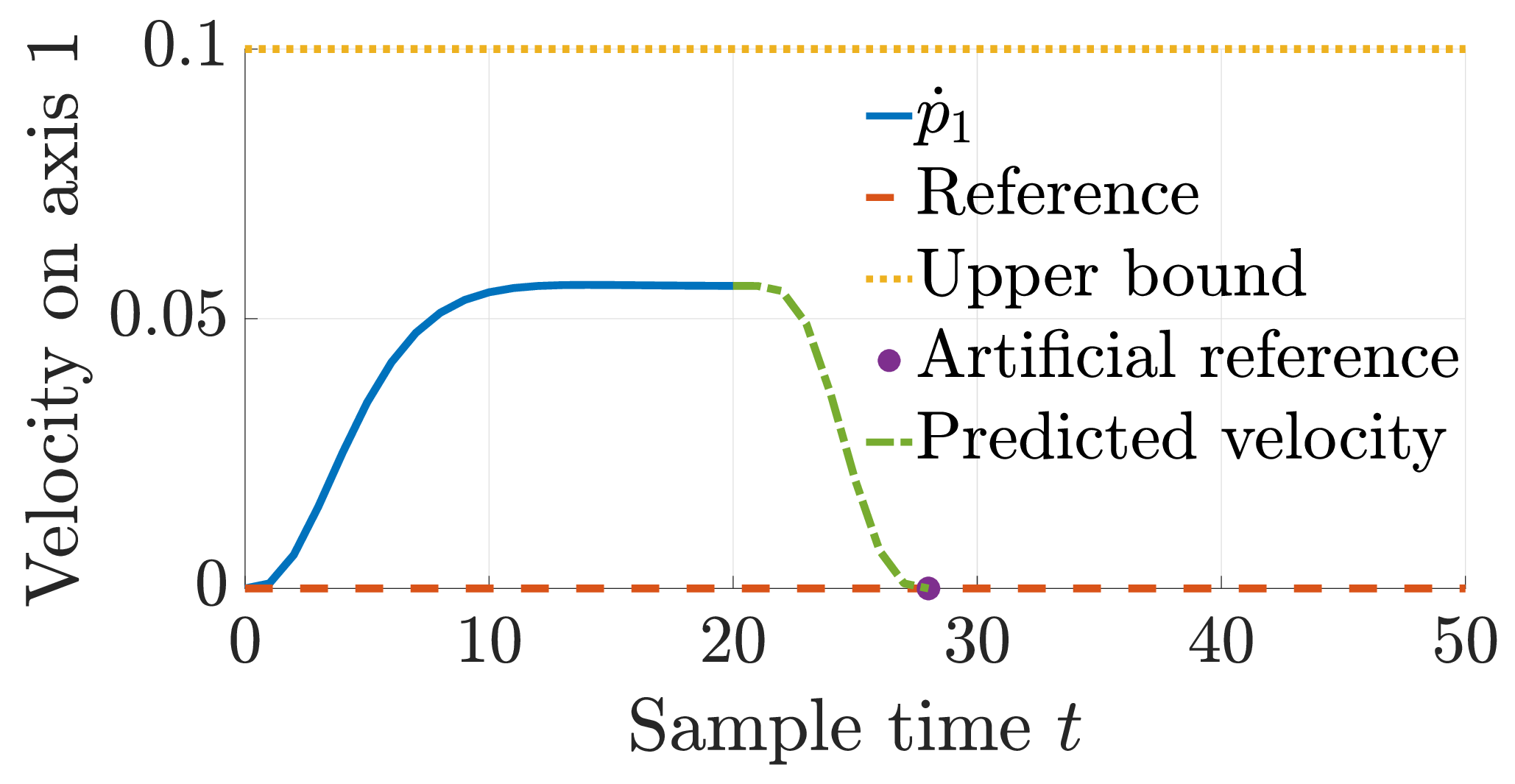}
        \caption{Sample time $20$ using MPCT~\eqref{eq:MPCT} with a prediction horizon of $N = 8$.}
        \label{fig:HMPC:example:iter}
    \end{subfigure}%

\caption{Example~\ref{example:HMPC}: performance issue of MPCT with small prediction horizons.}
    \label{fig:HMPC:example}
\end{figure}

\begin{example}[Performance of MPCT with small prediction horizons] \label{example:HMPC}
Let us consider the ball and plate system and MPCT controller presented in Section~\ref{sec:MPCT:BaP}.
Figure~\ref{fig:HMPC:example:vel} shows the closed-loop trajectory of the system using different prediction horizons, including the prediction horizon $N = 15$ used in Section~\ref{sec:MPCT:BaP}.
Note that the performance degrades significantly as the prediction horizon decreases, in that the convergence towards the desired reference is much slower.
The reason why this happens can be seen in Figure~\ref{fig:HMPC:example:iter}, which shows a snapshot of sample time $20$ of the simulation in Figure~\ref{fig:HMPC:example:vel} for the prediction horizon $N = 8$.
What we observe is that the velocity of the ball is not increasing much beyond the value of $0.05$, in spite of its upper bound being $0.1$.
The reason for this is the presence of the terminal equality constraints~\eqref{eq:MPCT:terminal} along with the other constraints of the system (in particular the input constraints).
It is easy to see that all steady states of the ball and plate system have a velocity equal to $0$, i.e., $\dot{p}_1 = \dot{p}_2 = 0$.
Thus, only states $x(t)$ that can reach a resting position in $N$ sample times may belong to the feasibility region of problem~\eqref{eq:MPCT}.
Therefore, the reason why the speed of the ball is so small in Figure~\ref{fig:HMPC:example:iter}, is that the velocity cannot be any higher due to the input constraints, or else there would be no feasible way of reaching a steady state in $N = 8$ sample times.
This issue is even more noticeable for smaller prediction horizons.
\end{example}

\begin{figure}[t]
    \centering
    \begin{subfigure}[ht]{0.48\textwidth}
        \includegraphics[width=\linewidth]{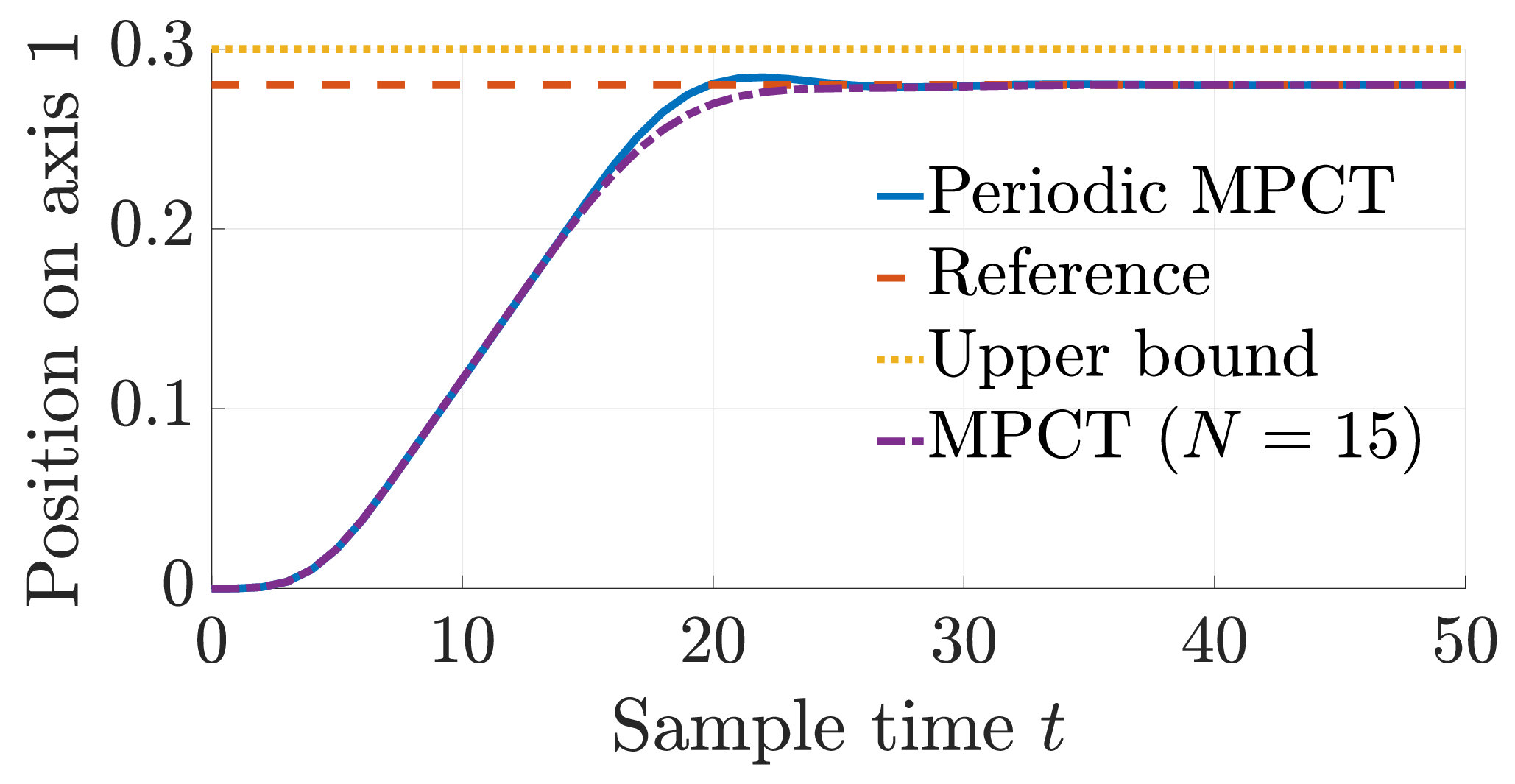}
        \caption{Trajectory of position of ball on axis $1$.}
        \label{fig:HMPC:example:perMPCT:pos}
    \end{subfigure}%
    \hfill%%
    \begin{subfigure}[ht]{0.48\textwidth}
        \includegraphics[width=\linewidth]{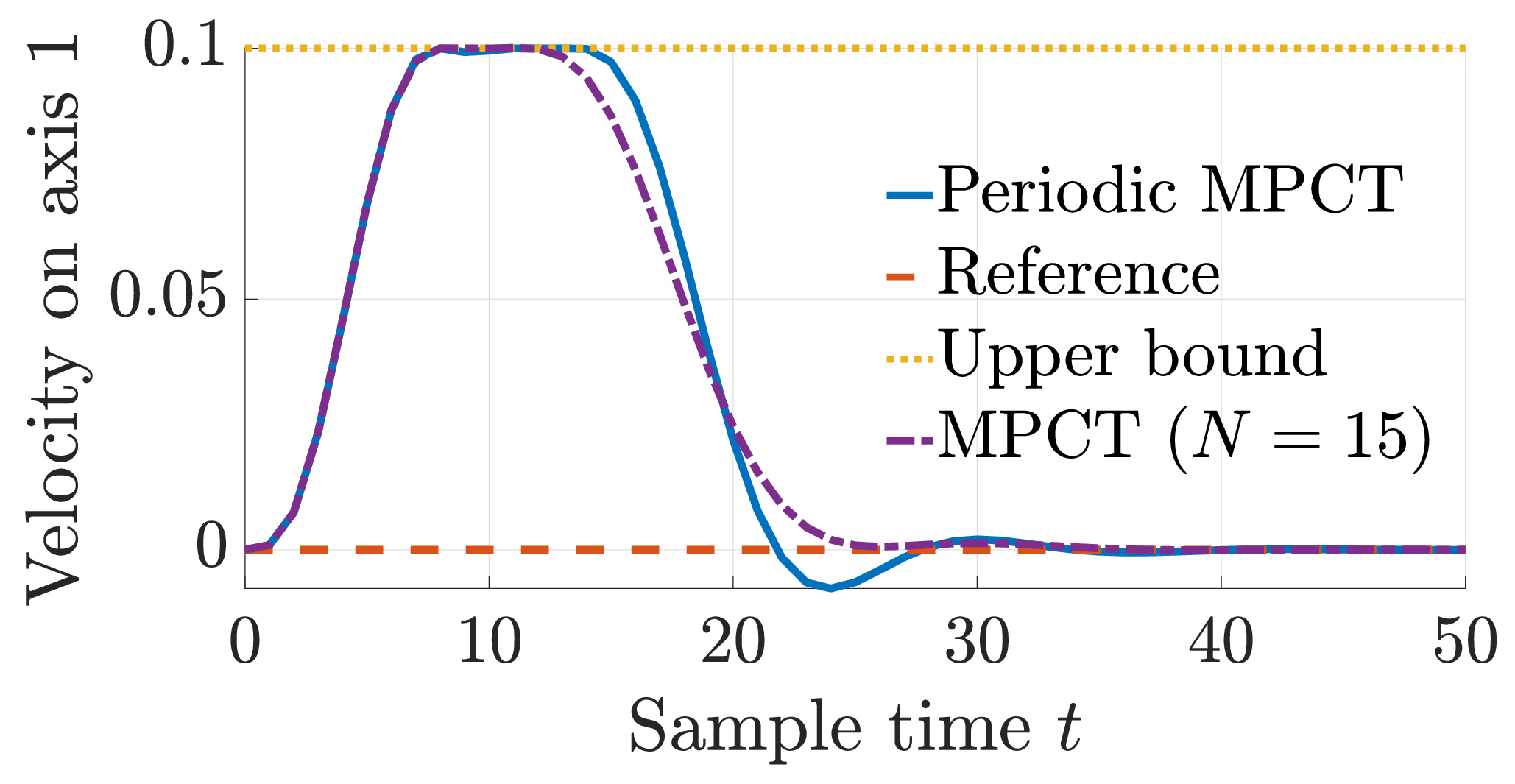}
        \caption{Trajectory of velocity of ball on axis $1$.}
        \label{fig:HMPC:example:perMPCT:vel}
    \end{subfigure}%

    \caption{Applying the periodic MPCT~\eqref{eq:perMPCT} with $N = 8$ and $\Tp = 15$ to Example~\ref{example:HMPC}.}
    \label{fig:HMPC:example:perMPCT}
\end{figure}

Obviously, the issue highlighted in Example~\ref{example:HMPC} becomes less prevalent as the prediction horizon is increased.
Indeed, after $N = 15$ there is no noticeable performance improvement\footnote{Which is the reason why this prediction horizon was chosen in the first place.}.
Still, it shows that the MPCT formulation can suffer from performance issues if the prediction horizon is too small.
In fact, we find that this issue is rather prevalent when dealing with systems with integral states and/or slew-rate constraints on the inputs\footnote{That is, constraints on the rate of change of the inputs.}.
The problem comes from the fact that the MPCT formulation~\eqref{eq:MPCT} requires the artificial reference $(\xs, \us)$ to be a steady state.
This, along with the system constraints~\eqref{eq:MPCT:constraints} and use of a terminal equality constraint~\eqref{eq:MPCT:terminal}, may restrict the feasibility region if the prediction horizon is too small.
Therefore, an immediate idea to solve this issue is to relax either the terminal equality constraint~\eqref{eq:MPCT:terminal} or the steady state requirement on the artificial reference~\eqref{eq:MPCT:art_ref:steady_state}.
The terminal equality constraint~\eqref{eq:MPCT:terminal} could be substituted by a more general terminal ingredient in the form of a classical terminal positive invariant set centered at $\xs$.
However, this would require the computation of a positive invariant set of the system and the addition of the constraints required to impose it in the optimization problem.

The other solution is to relax the steady state requirement of the artificial reference.
A first idea would be to use the periodic MPCT formulation~\eqref{eq:perMPCT}, since a steady state reference $(\xr, \ur)$ can be interpreted as a \emph{constant} periodic reference $(\xr(t), \ur(t))$ with any period $\Tp$.
Figure~\ref{fig:HMPC:example:perMPCT} shows the application of the periodic MPCT formulation~\eqref{eq:perMPCT} with $N = 8$ and $\Tp = 15$ to track the steady state of Figure~\ref{fig:HMPC:example}.
The results show that the periodic MPCT formulation with $N = 8$ has a similar performance to the MPCT formulation~\eqref{eq:MPCT} with a prediction horizon $N = 15$, cf. Figure~\ref{fig:HMPC:example:vel}.
Figure~\ref{fig:HMPC:example:perMPCT:vel} shows that in this case the velocity of the ball reaches its upper bound.
However, the issue with this solution is that, as discussed in Section~\ref{sec:MPCT:periodic}, the complexity of the periodic MPCT formulation~\eqref{eq:perMPCT} increases with the size of the period $\Tp$.
Thus, the use of the periodic MPCT formulation, as an alternative to MPCT~\eqref{eq:MPCT}, has little-to-no advantage over simply increasing the prediction horizon $N$ of~\eqref{eq:MPCT}.
Instead, we are interested in a tracking MPCT formulation that has a larger domain of attraction and better performance than the MPCT~\eqref{eq:MPCT}, when working with small prediction horizons, but without requiring a significantly larger computational cost.
This objective is achieved by the HMPC formulation.

\subsection{The harmonic MPC formulation} \label{sec:MPCT:HMPC}

Let us start by formally defining a \emph{harmonic signal} evolving in discrete time $t$.

\begin{definition}[Harmonic signal] \label{def:harmonic}
A discrete-time trajectory $v(t) \in \R^{m}$, with $t \in \N$, is a \emph{harmonic signal} if it satisfies
\begin{equation} \label{eq:harmonic}
    v(t) = v_e + v_s \sin(w t) + v_c \cos(w t), \\
\end{equation}
for some parameters $v_e, v_s, v_c \in \R^{m}$ and \emph{base frequency} $w > 0$.
\end{definition}

A harmonic signal $v(t)$ is determined by three parameters: $v_e$, which determines the \emph{bias} (or \emph{center}) of the signal, and $v_s$, $v_c$, which determine the magnitude of its \emph{sine} and \emph{cosine} terms, respectively.
Additionally, the period of the harmonic signal is determined by the choice of its \emph{base frequency} $w$.
For convenience, in the following we define the Cartesian product of the three parameters describing a harmonic signal $v(t)$ using the bold font $\vv{v}$, i.e., $\vv{v} \doteq (v_e, v_s, v_c) \in \R^m \times \R^m \times \R^m$.

The idea behind the HMPC formulation is to use a harmonic signal as its artificial reference.
Recall that the artificial references used in the previous sections are required to satisfy the system dynamics~\eqref{eq:model} and constraints~\eqref{eq:model:constraints}.
The same applies to the artificial reference of the HMPC formulation.
Therefore, before proceeding with the formulation itself, we first present some additional definitions and properties related to the satisfaction of system dynamics and constraint by a harmonic signal.

\begin{definition}[Admissible harmonic signals] \label{def:harmonic:admissible}
    The harmonic signals $\hat{x}(t)$ and $\hat{u}(t)$ with frequency $w > 0$, parametrized by $\vv{\hat{x}} \doteq (\hat{x}_e, \hat{x}_s, \hat{x}_c)$ and $\vv{\hat{u}} \doteq (\hat{u}_e, \hat{u}_s, \hat{u}_c)$, are said to be \emph{admissible} for system \eqref{eq:model} subject to \eqref{eq:model:constraints} if they satisfy ${\hat{x}(t+1)} = A \hat{x}(t) + B \hat{u}(t)$ and $\yLB \leq E \hat{x}(t) + F \hat{u}(t) \leq \yUB$, $\forall t \in \N$.
    Furthermore, we say that they are \emph{strictly admissible} if they satisfy $\yLB < E \hat{x}(t) + F \hat{u}(t) < \yUB$, $\forall t \in \N$.
\end{definition}

The following two propositions provide sufficient conditions for a harmonic signal to be admissible.
The reader can find their proofs in \cite{Krupa_TAC_2022}.
For clarity of presentation, in the following we use the notation
\begin{equation*}
    \hat{y}_e \doteq E \hat{x}_e + F \hat{u}_e, \; \hat{y}_s \doteq E \hat{x}_s + F \hat{u}_s, \; \hat{y}_c \doteq E \hat{x}_c + F \hat{u}_c.
\end{equation*}

\begin{proposition}[Satisfaction of system dynamics] \label{prop:harmonic:dynamics}
Let $\hat{x}(t)$ and $\hat{u}(t)$ be harmonic signals with the same frequency $w$ parametrized by $ \vv{\hat{x}} \doteq (\hat{x}_e, \hat{x}_s, \hat{x}_c)$ and $\vv{\hat{u}} \doteq (\hat{u}_e, \hat{u}_s, \hat{u}_c)$.
Consider the set
\begin{equation*}
    \cD \doteq \set{(\vv{\hat{x}}, \vv{\hat{u}})}{ \begin{array}{@{}l@{}} \hat{x}_e = A \hat{x}_e + B \hat{u}_e \\ \hat{x}_s \cos(w) - \hat{x}_c \sin(w) = A \hat{x}_s + B \hat{u}_s\\ \hat{x}_s \sin(w) + \hat{x}_c \cos(w) = A \hat{x}_c + B \hat{u}_c \end{array}}.
\end{equation*}
Then, $(\vv{\hat{x}}, \vv{\hat{u}}) \in \cD$ implies $\hat{x}(t+1) = A \hat{x}(t) + B \hat{u}(t)$, $\forall t \in \N$.
\end{proposition}

\begin{proposition}[Satisfaction of the system constraints] \label{prop:harmonic:constraints}
Let $\hat{x}(t)$ and $\hat{u}(t)$ be harmonic signals with the same frequency $w$ parametrized by $ \vv{\hat{x}} \doteq (\hat{x}_e, \hat{x}_s, \hat{x}_c)$ and $\vv{\hat{u}} \doteq (\hat{u}_e, \hat{u}_s, \hat{u}_c)$.
For $\sigma \geq 0$, consider the set
    \begin{equation*}
        \cC \doteq \set{(\hat{x}, \hat{u})}{ \begin{array}{@{}l@{}} (\hat{y}_{e(i)}, \hat{y}_{s(i)}, \hat{y}_{c(i)}) \in \overline{\cc{Y}}_i \cap \underline{\cc{Y}}_i, \, i \in \N_1^{n_y} \end{array}},
    \end{equation*}
where sets $\overline{\cc{Y}}_i$ and $\underline{\cc{Y}}_i$ are defined as
\begin{equation*} \label{eq:Y:sets}
\begin{aligned}
    \overline{\cc{Y}}_i &= \set{y = (y_0, y_1) \in \R\times\R^{2}}{\| y_1\| \leq \yUBj - \sigma - y_0}, \\
    \underline{\cc{Y}}_i &= \set{y = (y_0, y_1) \in \R\times\R^{2}}{\| y_1\| \leq y_0 - \yLBj - \sigma }.
\end{aligned}
\end{equation*}
Then, $(\vv{\hat{x}}, \vv{\hat{u}}) \in \cC$ implies $\yLB \leq E \hat{x}(t) + F \hat{u}(t) \leq \yUB$, $\forall t \in \N$.
Furthermore, the constraints are strictly satisfied if $\sigma > 0$.
\end{proposition}

\begin{corollary}
A harmonic signal satisfying Propositions~\ref{prop:harmonic:dynamics} and \ref{prop:harmonic:constraints} is an admissible harmonic signal of system \eqref{eq:model} subject to the constraints \eqref{eq:model:constraints}.
Furthermore, it is \emph{strictly} admissible if Propositions~\ref{prop:harmonic:constraints} is satisfied for some $\sigma > 0$.
\end{corollary}

The artificial reference of HMPC is taken as the harmonic signals
\begin{subequations} \label{eq:HMPC:art_ref}
\begin{align}
    \xh(k) &= \xe + \xs \sin(w k) + \xc \cos(w k), \label{eq:HMPC:x_h}\\
    \uh(k) &= \ue + \us \sin(w k) + \uc \cos(w k), \label{eq:HMPC:u_h}
\end{align}
\end{subequations}
with $\xh \in \R^\nx$, $\uh \in \R^\nu$, and where $\xH \doteq (\xe, \xs, \xc)$ and $\uH \doteq (\ue, \us, \uc)$ are decision variables of the HMPC optimization problem:
\begin{subequations} \label{eq:HMPC} % HMPC
\begin{align}
    \min\limits_{\vv{x},\vv{u}, \xH, \uH} \;& \Sum{k=0}{N-1} \left( \| x(k|t) - \xh(k) \|_Q^2 + \| u(k|t) - \uh(k) \|_R^2 \right) + \VfHMPC(\xH, \uH; \xr, \ur) \label{eq:HMPC:cost}\\
             \st \;& x(k+1|t) = A x(k|t) + B u(k|t), \; k \in \N_0^{N-1}, \label{eq:HMPC:dynamics}\\
                  & \yLB \leq E x(k|t) + F u(k|t) \leq \yUB, \; k \in \N_0^{N-1}, \label{eq:HMPC:constraints}\\
                  & x(0|t) = x(t), \label{eq:HMPC:x0}\\
                  & x(N|t) = \xe + \xs \sin(w N) + \xc \cos(w N), \label{eq:HMPC:xN}\\
                  & (\xH, \uH) \in \cD, \label{eq:HMPC:D} \\
                  & (\xH, \uH) \in \cC, \label{eq:HMPC:C}
\end{align}
\end{subequations}
with offset cost function
\begin{equation} \label{eq:HMPC:offsetCost}
    \VfHMPC(\xH, \uH; \xr, \ur) = \| \xe - \xr \|_{T_e}^2 + \| \ue - \ur \|_{S_e}^2 + \| \xs \|_{T_h}^2 + \| \xc \|_{T_h}^2 + \| \us \|_{S_h}^2 + \| \uc \|_{S_h}^2,
\end{equation}
and where $\vv{x} = (x(0|t), \dots, x(N|t) )$, $\vv{u} = ( u(0|t), \dots, u(N-1|t) )$ and $\sigma \geq 0$ plays the same role as in the MPCT formulation~\eqref{eq:MPCT}.
We assume that $Q$, $R$, $T_e$ and $S_e$ are positive definite matrices and that $T_h$ and $S_h$ are diagonal positive definite matrices.

Constraints \eqref{eq:HMPC:dynamics}, \eqref{eq:HMPC:constraints} and \eqref{eq:HMPC:x0} impose the typical MPC constraints; namely the imposition of the system dynamics, system constraints and current system state.
Constraint \eqref{eq:HMPC:xN} is imposing the value of the terminal state $x(N|t)$ to be equal to $\xh(N)$, c.f., \eqref{eq:HMPC:art_ref}.
That is, the terminal state is required to reach the artificial reference, as in MPCT~\eqref{eq:MPCT} and periodic MPCT~\eqref{eq:perMPCT}.
Finally, \eqref{eq:HMPC:D} and \eqref{eq:HMPC:C} are imposing that the artificial harmonic reference satisfies the system dynamics and constraints, as shown in Propositions~\ref{prop:harmonic:dynamics} and~\ref{prop:harmonic:constraints}.

Note that the design parameter $w$ of the HMPC formulation~\eqref{eq:HMPC} determines the period of its artificial harmonic reference $(\xh, \uh)$.
However, the complexity of optimization problem~\eqref{eq:HMPC} is independent of the value of $w$, since the number of constraints~\eqref{eq:HMPC:D} and~\eqref{eq:HMPC:C} required to impose the system dynamics and constraints on the artificial harmonic reference is fixed.
Thus, parameter $w$ can be freely chosen to increase the performance of the closed-loop system controlled with the HMPC formulation.
This provides the main benefit of HMPC w.r.t. using the periodic MPCT formulation~\eqref{eq:perMPCT} to track a steady state desired reference.
A detailed discussion of how to choose $w$ is not included here due to space considerations.
Instead, the reader is referred to~\cite[\S VI.B]{Krupa_TAC_2022} for discussion on how to select it.

Note that the stage cost in~\eqref{eq:HMPC:cost} is no different from the one used in the tracking MPC formulations presented in previous sections.
That is, it penalizes the discrepancy between the predicted states and inputs with the value of the artificial reference along the prediction horizon.
On the other hand, the offset cost function is conceptually different to the ones used in the previous MPC for tracking formulations, although the overall idea is the same. 
The offset cost \eqref{eq:HMPC:offsetCost} penalizes, on one hand, the discrepancy between the \emph{center} $(\xe, \ue)$ of the artificial harmonic reference \eqref{eq:HMPC:art_ref} with the desired reference $(\xr, \ur)$, and on the other, the magnitude of the sine and cosine terms $\xs, \xc, \us, \uc$ of the artificial harmonic reference.
The end result of this is that as time $t$ increases, $(\xe(t), \ue(t))$ will converge towards $(\xr, \ur)$ and $\xs(t), \xc(t), \us(t), \uc(t)$ will all converge to $0$.
The latter is always true, as we will formally state in what follows.
That is, $\xs(t), \xc(t), \us(t), \uc(t)$ all asymptotically converge to $0$ as $t \to \infty$.
The former, however, is only true if $(\xr, \ur)$ is an admissible steady state of the system.
Otherwise, as in the classical MPCT formulation~\eqref{eq:MPCT}, the HMPC formulation will converge to the admissible steady state that is closest to $(\xr, \ur)$, as measured by the offset cost function~\eqref{eq:HMPC:offsetCost}.
Additionally, HMPC also guarantees recursive feasibility even if the desired reference is changed online.
The following two theorems formalize the recursive feasibility and asymptotic stability properties of the HMPC formulation.
Their proofs can be found in~\cite{Krupa_TAC_2022}.

\begin{theorem}[Recursive feasibility of the HMPC formulation] \label{theo:HMPC:feasibility}
Let $x(t)$ belong to the feasibility region of the HMPC formulation~\eqref{eq:HMPC}.
Let $\vv{x}$, $\vv{u}$, $\xH$, $\uH$ be any feasible solution of~\eqref{eq:HMPC} for a given reference $(\xr(t), \ur(t))$.
Then, the successor state $A x(t) + B \tilde{u}(0|t)$ belongs to the feasibility region of~\eqref{eq:HMPC} for any reference $(\xr(t+1), \ur(t+1))$.
\end{theorem}

The HMPC formulation asymptotically stabilizes the system to the \emph{optimal reachable reference} determined by its offset cost function.

\begin{definition}[Optimal reachable reference of HMPC~\eqref{eq:HMPC}]
Given a reference $(\xr, \ur) \in \R^\nx \times \R^\nu$, we define the \emph{optimal reachable reference} of the HMPC formulation~\eqref{eq:HMPC} as the harmonic sequences $\{\xh^\circ(k)\}$, $\{\uh^\circ(k)\}$, $k \in \N$, parameterized by the unique solution $(\xHo, \uHo)$ of the strongly convex optimization problem
\begin{equation} \label{eq:HMPC:OP:optimal:harmonic:refrefence}
\begin{aligned}
    (\xHo, \uHo) = \arg\min\limits_{\xH, \uH} \;& \VfHMPC(\xH, \uH; \xr, \ur) \\
                    \st\;&  (\xH, \uH) \in \cD, \\
                         & (\xH, \uH) \in \cC.
\end{aligned}
\end{equation}
\end{definition}

The following proposition shows that, even though the HMPC formulation uses a \emph{harmonic} artificial reference, its optimal reachable reference is always given by a admissible steady state of the system.
Its proof can also be found in~\cite{Krupa_TAC_2022}.

\begin{proposition}[Characterization of $(\xHo, \uHo)$ for the HMPC formulation \eqref{eq:HMPC}] \label{prop:HMPC:optimal:harmonic:reference}
Consider the HMPC formulation~\eqref{eq:HMPC}.
Then, for any $(\xr, \ur) \in \R^\nx \times \R^\nu$, the optimal solution $(\xHo, \uHo)$ of problem~\eqref{eq:HMPC:OP:optimal:harmonic:refrefence} is given by $\xHo = (\xeo, 0, 0)$ and $\uHo = (\ueo, 0, 0)$, where $(\xeo, \ueo) \in \R^\nx \times \R^\nu$ is the steady state of \eqref{eq:model} satisfying 
\begin{equation} \label{eq:HMPC:optimal:harmonic:reference:constraints}
    \yLB + \sigma \ones{\ny} \leq E \xeo + F \ueo \leq \yUB - \sigma \ones{\ny}
\end{equation}
that minimizes the cost $\| \xeo - \xr \|^2_{T_e} + \| \ueo - \ur \|^2_{S_e}$.
\end{proposition}

\begin{remark}
    The optimal reachable references of the HMPC formulation \eqref{eq:HMPC} and the MPCT formulation \eqref{eq:MPCT} coincide if $T_e = T$, $S_e = S$ and both formulations consider the same $\sigma$, cf. Proposition~\ref{prop:HMPC:optimal:harmonic:reference} and Definition~\ref{def:MPCT:orr}.
\end{remark}

\begin{theorem}[Asymptotic stability of the HMPC formulation] \label{theo:HMPC:stability}
    Let $(\xr, \ur)$ be a fixed desired reference.
    Assume that the initial state $x(0)$ belongs to the feasibility region of~\eqref{eq:HMPC} and that the prediction horizon $N$ is greater or equal to the controllability index of system~\eqref{eq:model}.
    Then, system~\eqref{eq:model} controlled with the control law of the HMPC formulation~\eqref{eq:HMPC} is stable, fulfills the system constraints for all $t \in \N$, and asymptotically converges to the optimal reachable reference $(\xeo, \ueo) \in \R^\nx \times \R^\nu$ given by Proposition~\ref{prop:HMPC:optimal:harmonic:reference}.
\end{theorem}

The user may identify an apparent downside of the HMPC formulation~\eqref{eq:HMPC}: its optimization problem~\eqref{eq:HMPC} is not a QP problem, as is the case in the MPCT formulation~\eqref{eq:MPCT}.
This is due to the inclusion of the constraints~\eqref{eq:HMPC:C}, which are second order cone (SOC) constraints.
Therefore, problem~\eqref{eq:HMPC} is a SOC programming problem, which in general is a class of optimization problem that is more difficult to solve than the well-studied class of QP problems.
However, there are several efficient state-of-the-art solvers for SOC programming problems in the literature~\cite{ODonoghue_SCS_21,Garstka_JOTA_2021}. 
Furthermore, in~\cite{Krupa_TAC_2023} the authors presented a solver for the HMPC formulation that is designed to efficiently deal with the SOC constraints~\eqref{eq:HMPC:C}.
The results in~\cite{Krupa_TAC_2023} show that the HMPC formulation~\eqref{eq:HMPC} can be solved in computation times comparable to solving the MPCT formulation~\eqref{eq:MPCT} using state-of-the-art QP solvers.

\begin{remark}
As a final remark, the authors would like to point out that the HMPC formulation~\eqref{eq:HMPC} has been recently extended in~\cite{Krupa_arXiv_ellipHMPC_23} to the problem of tracking harmonic reference trajectories~\eqref{eq:harmonic}.
\end{remark}

\subsection{Performance benefits of the HMPC formulation: case study on the ball and plate system} \label{sec:HMPC:results}

\begin{figure}[t]
    \centering
    \begin{subfigure}[ht]{0.48\textwidth}
        \includegraphics[width=\linewidth]{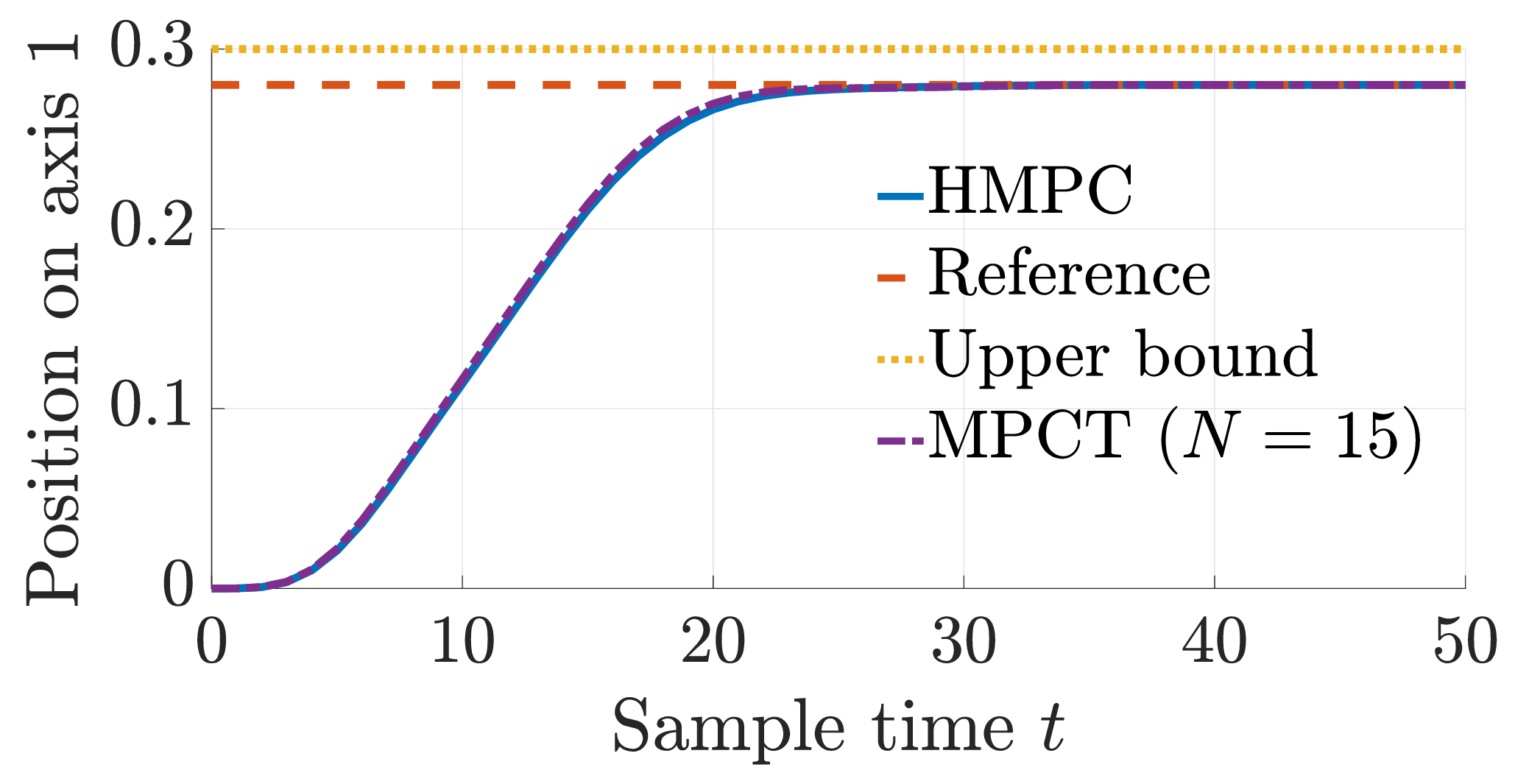}
        \caption{Trajectory of position of ball on axis $1$.}
        \label{fig:HMPC:example:HMPC:vel}
    \end{subfigure}%
    \hfill%%
    \begin{subfigure}[ht]{0.48\textwidth}
        \includegraphics[width=\linewidth]{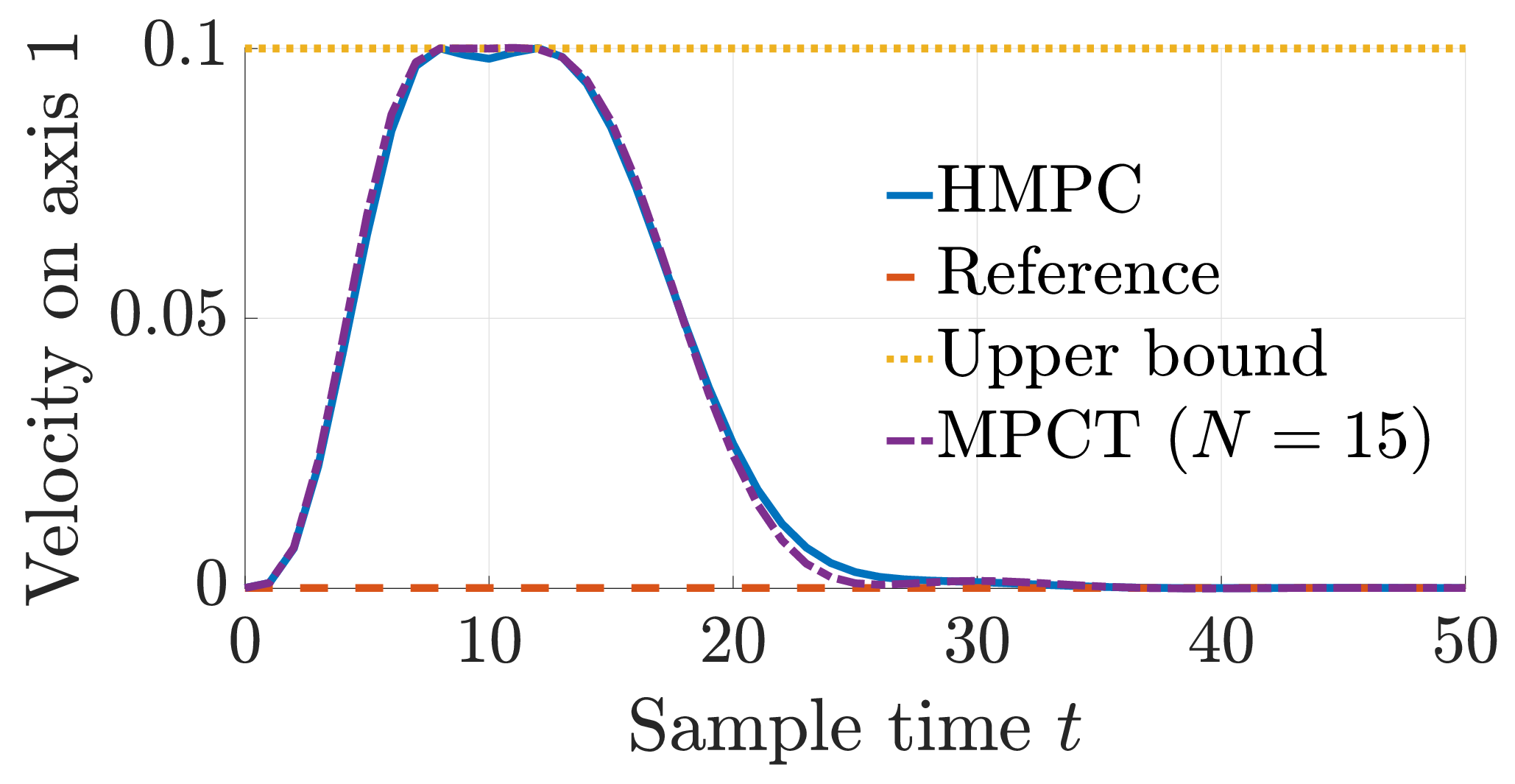}
        \caption{Trajectory of velocity of ball on axis $1$.}
        \label{fig:HMPC:example:HMPC:iter}
    \end{subfigure}%

    \caption{Applying HMPC~\eqref{eq:HMPC} with $N = 8$ to Example~\ref{example:HMPC}.}
    \label{fig:HMPC:example:HMPC}
\end{figure}

As discussed in the introduction of this section, one of the main benefits of HMPC is its increased performance w.r.t. the MPCT formulation~\eqref{eq:MPCT} when working with small prediction horizons.
Indeed, Figure~\ref{fig:HMPC:example:HMPC} shows the application of the HMPC formulation~\eqref{eq:HMPC} to Example~\ref{example:HMPC}.
As seen in Figure~\ref{fig:HMPC:example:HMPC:vel}, HMPC with $N = 8$ has a very similar behavior (and performance) to MPCT with $N = 15$.
To see why, compare Figures~\ref{fig:HMPC:example:iter} and~\ref{fig:HMPC:example:HMPC:iter}.
As seen in Figure~\ref{fig:HMPC:example:HMPC:iter}, the speed of the ball reaches its maximum bound when using the HMPC formulation, even though the prediction horizon is $N = 8$.
The reason why this is possible is that HMPC does not require the terminal state $x(N|t)$ to reach a steady state of the system.
Instead, it requires it to reach some harmonic trajectory of the system.
In terms of the position of the ball on the plate, this translates into a periodic trajectory of the ball on its surface.

Regarding the benefits of HMPC in terms of domain of attraction, we refer the reader to~\cite{Krupa_CDC_19} for some numerical results highlighting the difference between MPCT~\eqref{eq:MPCT} and HMPC~\eqref{eq:HMPC} on an academic example.
The argument for the larger domain of attraction of HMPC is simple: it is a generalization of the MPCT formulation~\eqref{eq:MPCT}\footnote{Note that HMPC~\eqref{eq:HMPC} is equivalent to MPCT~\eqref{eq:MPCT} if $w = 0$ or any multiple of $2 \pi$.} with more degrees of freedom.
In other words, the terminal state $x(N|t)$ only needs to be able to reach \emph{any} admissible harmonic signal of the system, instead of a steady state.

\begin{figure}[t]
    \centering
    \begin{subfigure}[ht]{0.49\textwidth}
        \includegraphics[width=\linewidth]{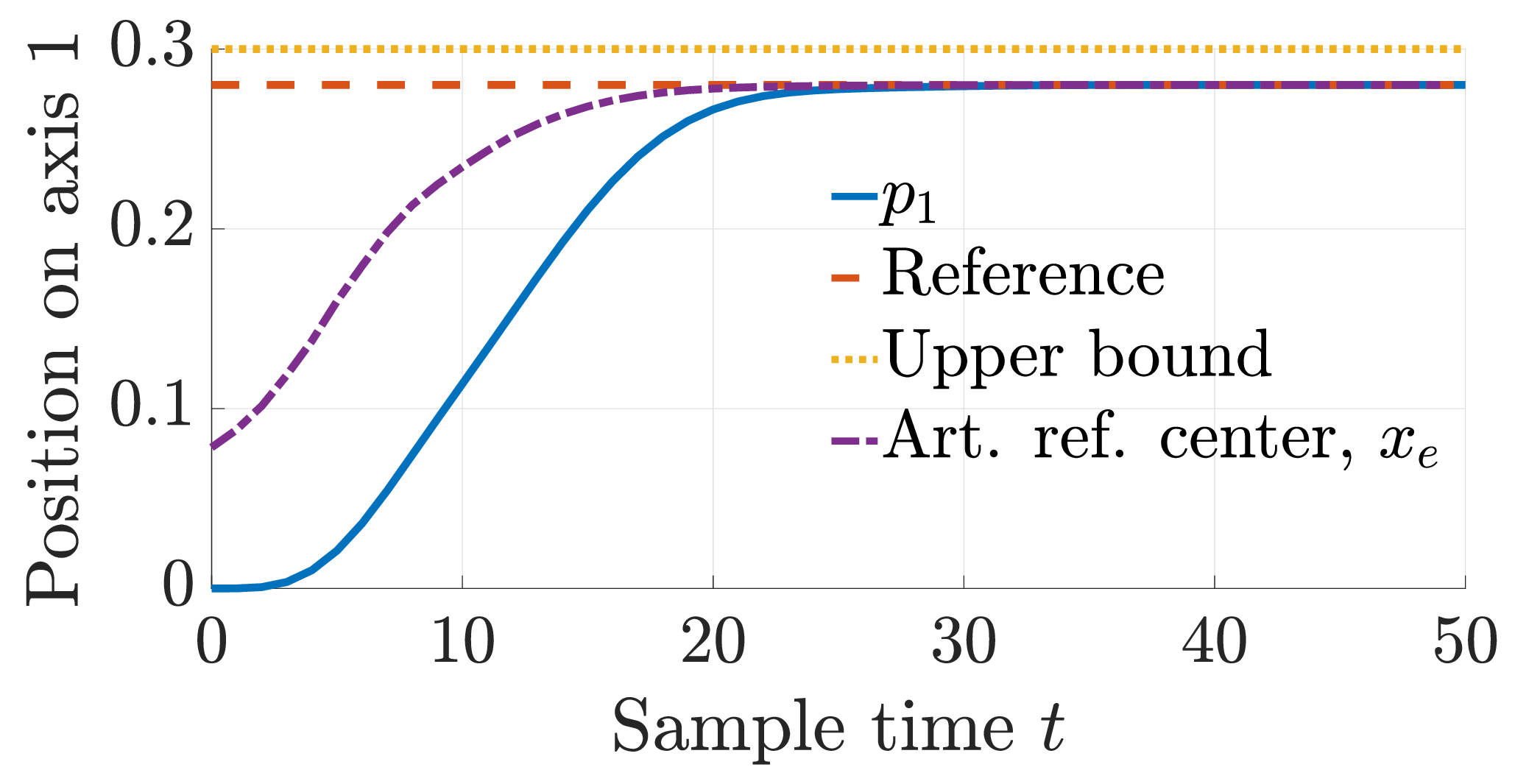}
        \caption{Trajectory for admissible reference.}
        \label{fig:BaP:HMPC:ad:traj}
    \end{subfigure}%
    \hfill%%
    \begin{subfigure}[ht]{0.49\textwidth}
        \includegraphics[width=\linewidth]{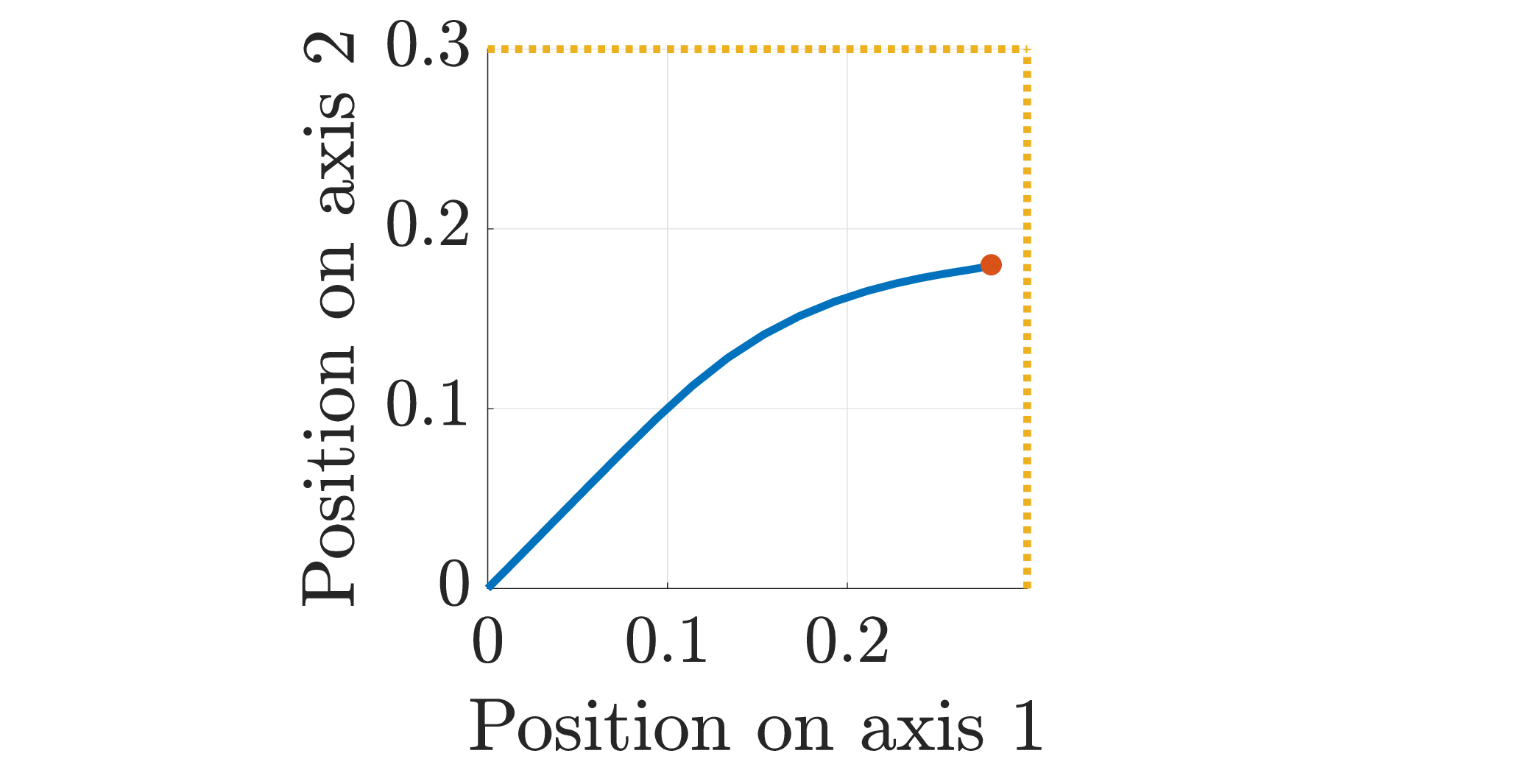}
        \caption{Position of ball for admissible reference.}
        \label{fig:BaP:HMPC:ad:pos}
    \end{subfigure}%

    \begin{subfigure}[ht]{0.49\textwidth}
        \includegraphics[width=\linewidth]{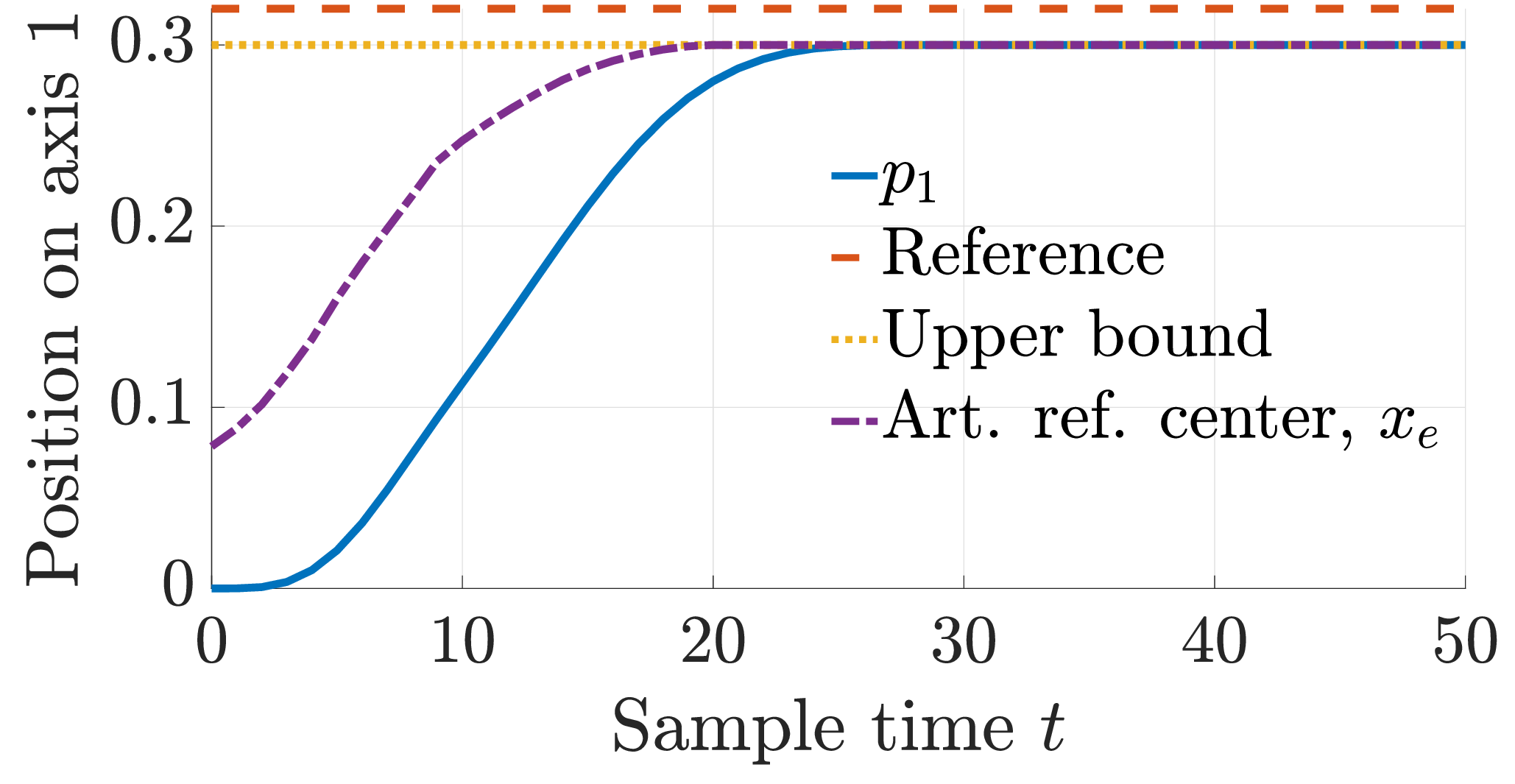}
        \caption{Trajectory for non-admissible reference.}
        \label{fig:BaP:HMPC:nonad:traj}
    \end{subfigure}%
    \hfill%%
    \begin{subfigure}[ht]{0.49\textwidth}
        \includegraphics[width=\linewidth]{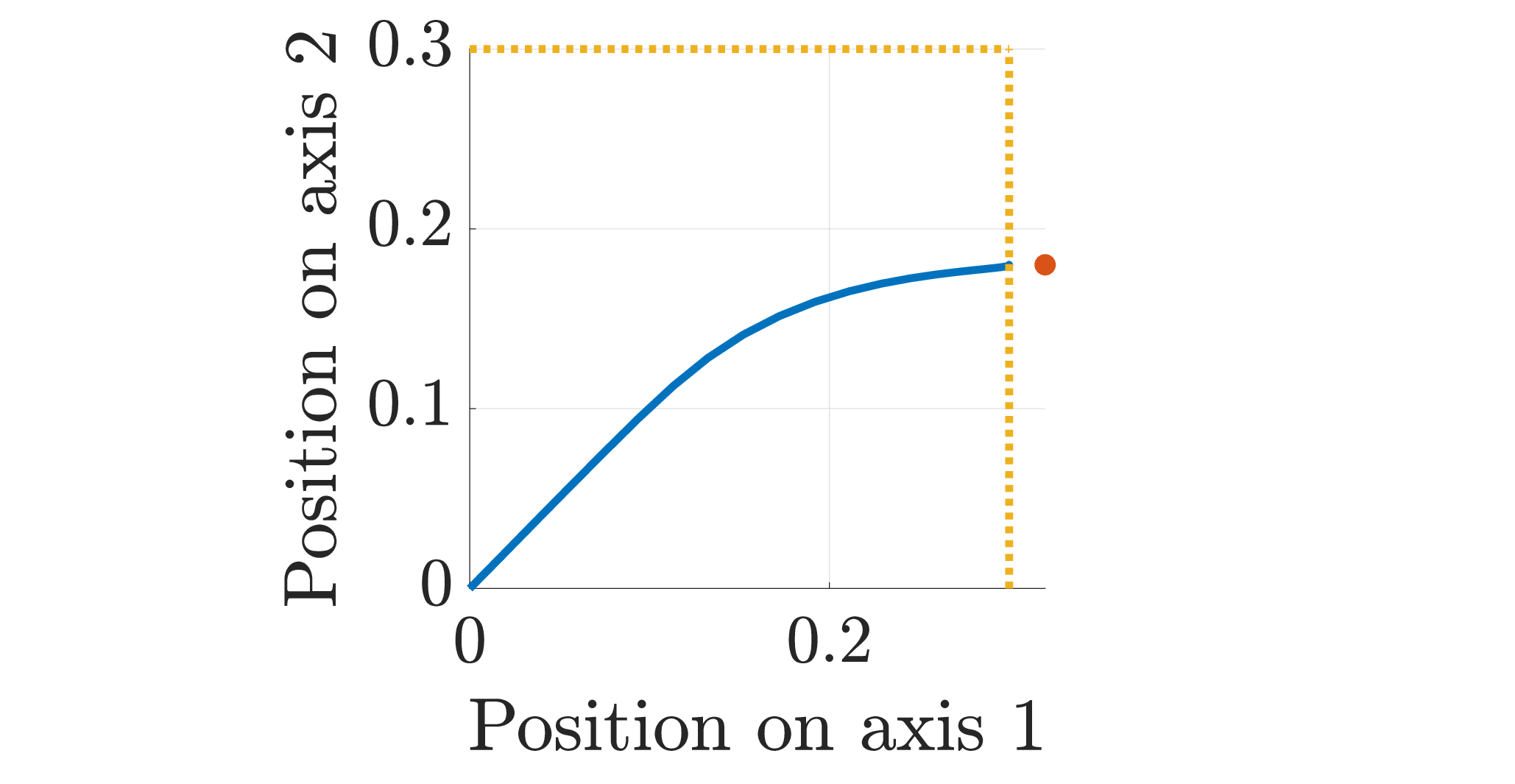}
        \caption{Position of ball for non-admissible reference.}
        \label{fig:BaP:HMPC:nonad:pos}
    \end{subfigure}%

    \caption{Closed-loop results of the ball and plate system using HMPC~\eqref{eq:HMPC}.}
    \label{fig:BaP:HMPC}
\end{figure}

Finally, Figure~\ref{fig:BaP:HMPC} shows closed-loop results on the ball and plate system presented in Section~\ref{sec:MPCT:BaP} using the HMPC formulation~\eqref{eq:HMPC}.
The figure considers the same setup and steady state references used in Section~\ref{sec:MPCT:BaP} (cf. Figure~\ref{fig:BaP:MPCT}), taking the HMPC ingredients as $N = 8$, $Q = \diag(10, 0.05, 0.05, 0.05, 10, 0.05, 0.05, 0.05)$, $R = \diag(0.5, 0.5)$, $T_e = N Q$, $S_e = N R$, $T_h = T_e$, $S_h = S_e$ and $w = 0.3254$.
Note that we take the same $N$, $Q$, $R$ used in Section~\ref{sec:MPCT:BaP} for the MPCT formulation~\eqref{eq:MPCT}.
Furthermore, $T_e$ and $S_e$ are taken as the ingredients $T$ and $S$ used for the MPCT formulation~\eqref{eq:MPCT} in Section~\ref{sec:MPCT:BaP}.
Figures~\ref{fig:BaP:HMPC:ad:traj} and~\ref{fig:BaP:HMPC:nonad:traj} show the trajectory of the element of $x_e$ corresponding to the position $p_1$ of the ball.
Note that its evolution is very similar to the artificial reference $x_s$ of the MPCT formulation shown in Figure~\ref{fig:BaP:MPCT}.
In fact, the closed-loop results shown in Figure~\ref{fig:BaP:HMPC} are very similar to the ones obtained in Figure~\ref{fig:BaP:MPCT}, i.e., to the ones obtained using the MPCT formulation~\eqref{eq:MPCT} with the same ingredients but with a prediction horizon $N = 15$ (as also seen in Figure~\ref{fig:HMPC:example:HMPC}).
The results highlight the performance advantage of the HMPC formulation when working with small prediction horizons, since the MPCT formulation~\eqref{eq:MPCT} with a prediction horizon $N = 8$ performs much worse, as shown in Example~\ref{example:HMPC} and Figure~\ref{fig:HMPC:example}.
Finally, note that the HMPC formulation steers the system to a steady state, even though its artificial reference is a harmonic signal, as stated in Proposition~\ref{prop:HMPC:optimal:harmonic:reference}.

\begin{remark}
    The closed-loop results for the HMPC formulation are obtained using version \texttt{v0.3.11} of the Spcies toolbox for MATLAB~\cite{Spcies}.
    For comparisons between HMPC and MPCT in terms of their computation times, we refer the reader to~\cite{Krupa_TAC_2023}.
\end{remark}

\section{Concluding remarks} \label{sec:conclusions}

This chapter has presented a light introduction to the classical \emph{tracking} MPC formulation for piecewise-constant references, illustrating its benefits w.r.t. classical MPC formulations.
Then, it has presented its extension for tracking periodic references as well as a recent extension, known as \emph{harmonic} MPC, which draws inspiration from the periodic \emph{tracking} MPC formulation.
We have shown that \emph{harmonic} MPC is designed to provide a very good performance and a large domain of attraction when working with very small prediction horizons.

% Fakesection Acknowledgments
\begin{acknowledgement}
The authors acknowledge support from Grant PID2022-141159OB-I00 funded by MICIU/AEI/10.13039/501100011033 and by ERDF/EU;
Grant PDC2021-121120-C21 funded by MCIN/AEI/10.13039/501100011033 and by the “European Union NextGeneration EU/PRTR”;
the MUR-PRO3 project on Software Quality;
and the MUR-PRIN project DREAM (20228FT78M).
\end{acknowledgement}

% Fakesection Bibliography
\bibliographystyle{IEEEtran}
\bibliography{IEEEabrv,BibChapMPCT}

\end{document}